\documentclass[12pt]{oxarticle}
\pdfoutput=1
\usepackage{graphicx, float, array, xspace, amscd, amsmath, amsthm, amssymb, latexsym, relsize, bbm}
\usepackage{geometry}
\usepackage[dvipsnames]{xcolor}
\allowdisplaybreaks		%
\usepackage[bbgreekl]{mathbbol}
\usepackage{amsfonts, tikz-cd}
\PassOptionsToPackage{linecolor=blue,backgroundcolor=blue!25,bordercolor=blue,textsize=scriptsize}{todonotes}
\usepackage{setspace}
\usepackage{slashed}

\DeclareSymbolFontAlphabet{\mathbb}{AMSb}
\DeclareSymbolFontAlphabet{\mathbbl}{bbold}

\usepackage{cite}
\usepackage[all,cmtip]{xy}
\usepackage{color}
\definecolor{darkblue}{rgb}{0.0,0.0,0.5} 	%
\usepackage[linktocpage=true]{hyperref}
\hypersetup{colorlinks=true, citecolor=darkblue, linkcolor=darkblue, urlcolor=darkblue, pdfauthor={}, pdftitle={},breaklinks=true}
\usepackage{cancel}

\DeclareSymbolFont{rsfs}{U}{rsfs}{m}{n}
\DeclareSymbolFontAlphabet{\mathscrsfs}{rsfs}

\global\long\def\SU#1{\text{SU}(#1)}%
\global\long\def\Uni#1{\text{U}(#1)}%
\global\long\def\SL#1{\text{SL}(#1)}%
\global\long\def\op#1{\operatorname{#1}}%

\renewcommand{\k}{\kappa}

\newcommand{\Dc}{\mathcal{D}}

\DeclareFontFamily{OT1}{pzc}{}
\DeclareFontShape{OT1}{pzc}{m}{it}{<-> s * [1.200] pzcmi7t}{}
\DeclareMathAlphabet{\mathpzc}{OT1}{pzc}{m}{it}

\newcommand{\cN}{\mathcal{N}}

\newcommand{\bC}{\mathbb{C}}

\DeclareFontFamily{U}{bbold}{}
\DeclareFontShape{U}{bbold}{m}{n}
 {  <-5.5> s*[1.05] bbold5
    <5.5-6.5> s*[1.05] bbold6
    <6.5-7.5> s*[1.05] bbold7
    <7.5-8.5> s*[1.05] bbold8
    <8.5-9.5> s*[1.05] bbold9
    <9.5-11.5> s*[1.05] bbold10
    <11.5-16> s*[1.05] bbold12
    <16-> s*[1.05] bbold17
 }{}

\font\eightrm=cmr8 at 8pt

\newcommand{\beq}{\begin{equation}}
\newcommand{\eeq}{\end{equation}}
\newcommand{\beqnn}{\begin{equation*}}
\newcommand{\eeqnn}{\end{equation*}}
\newcommand{\bea}{\begin{eqnarray}}
\newcommand{\eea}{\end{eqnarray}}
\newcommand{\bean}{\begin{eqnarray*}}
\newcommand{\eean}{\end{eqnarray*}}

\newcommand{\ee}{\text{e}}
\newcommand{\ii}{\text{i}}

\newcommand{\place}[3]{\vbox to0pt{\kern-\parskip\kern-7pt
                             \kern-#2truein\hbox{\kern#1truein #3}
                             \vss}\nointerlineskip}

\DeclareFontFamily{U}{wncy}{}
\DeclareFontShape{U}{wncy}{m}{n}{<->wncyr10}{}
\DeclareSymbolFont{mcy}{U}{wncy}{m}{n}
\DeclareMathSymbol{\sha}{\mathord}{mcy}{"58}

\newcommand{\del}{{\partial}}
\newcommand{\delb}{{\bar{\partial}}}

\usepackage{tensor}

\newcommand{\dd}{{\text{d}}}

\newcommand{\Vol}{\operatorname{Vol}}
\newcommand{\tr}{\operatorname{tr}}
\newcommand{\Tr}{\operatorname{Tr}}
\newcommand{\rk}{\operatorname{rk}}
\newcommand{\End}{\operatorname{End}}

\makeatletter
\g@addto@macro\bfseries{\boldmath}
\makeatother

\global\long\def\OmAB{\overset{\bullet}{\Omega}}%
\global\long\def\OmCD{\underset{\bullet}{\Omega}}%
\global\long\def\OmAC{\raisebox{1pt}{$\scriptstyle\bullet$}\Omega}%
\global\long\def\OmBD{\Omega\raisebox{1pt}{$\scriptstyle\bullet$}}%

\global\long\def\smallbullet{\raisebox{1pt}{$\scriptstyle\bullet$}}%

\newcommand{\der}{\partial}
\newcommand{\Qhol}{\mathscrsfs{D}}

\newcommand{\FDet}{\mathcal{F}_{\mathrm{Det}}}

\newcommand{\AT}{\mathcal{T}}

\newcommand{\Q}{\mathcal{Q}} %

\makeatletter
\g@addto@macro\bfseries{\boldmath}
\makeatother
\let\originalleft\left
\let\originalright\right
\renewcommand{\left}{\mathopen{}\mathclose\bgroup\originalleft}
\renewcommand{\right}{\aftergroup\egroup\originalright}
\usepackage{microtype}
\usepackage{mathtools}
\mathtoolsset{centercolon}
\def\eqspace{\mathrel{\phantom{{=}}{}}}

\global\long\def\td{\operatorname{td}}%
\global\long\def\ch{\operatorname{ch}}%
\global\long\def\rk{\operatorname{rk}}%
\global\long\def\id{\boldsymbol{1}}%

\numberwithin{equation}{section}
\proofmodefalse
\begin{document}
\begin{titlepage}
\begin{flushright} MI-HET-806
\end{flushright}
\vfill
\begin{center}
{\huge A Heterotic Kodaira--Spencer Theory at One-Loop\par} 

\vskip 1cm 

{Anthony Ashmore$^+$, Javier Jos\'{e} Murgas Ibarra$^{\dagger}$, David Duncan McNutt$^{\times}$,\\
Charles Strickland-Constable$^\$$, Eirik Eik Svanes$^{\dagger}$, David Tennyson$^{*}$, Sander Winje$^{\dagger}$}

\vskip 1cm 

{\it 

$^+$Enrico Fermi Institute and Kadanoff Center for Theoretical Physics\\
University of Chicago, Chicago, IL 60637, USA\\[0.2cm]

$^\dagger$Department of Mathematics and Physics, Faculty of Science and Technology\\
University of Stavanger, N-4036, Stavanger, Norway\\[0.2cm]

$^{\times}$Department of Mathematics and Statistics \\
UiT: The Arctic University of Norway, N-9019, Tromso, Norway\\[0.2cm]

$^*$Mitchell Institute for Fundamental Physics and Astronomy\\
Texas A\&M University, College Station, TX, 77843, USA\\[0.2cm]

$^\$$Department of Physics, Astronomy and Mathematics, \\
University of Hertfordshire, College Lane, Hatfield, AL10 9AB, United Kingdom
}  
\end{center}
\vfill
\begin{center}
    {\bf Abstract}
\end{center}
\begin{quote}
We consider a heterotic version of six-dimensional Kodaira--Spencer gravity derived from the heterotic superpotential. We compute the one-loop partition function and find it can be expressed as a product of holomorphic Ray--Singer torsions. We discuss its topological properties and potential gauge and gravitational anomalies. We show these anomalies can be cancelled using Green--Schwarz-like counter-terms. We also discuss the dependence on the background geometry, and in particular the choice of hermitian metric needed for quantisation. Given suitable topological constraints, this dependence may again be cancelled by the addition of purely background-dependent counter-terms. 
We also explain how our methods provide the one-loop partition functions of a large class of more general holomorphic field theories in terms of holomorphic Ray--Singer torsions.
\end{quote}
\vfill

\end{titlepage}

\newpage
\microtypesetup{protrusion=false} %
\tableofcontents %
\microtypesetup{protrusion=true} %
\setcounter{page}{1}
\pagestyle{plain}

\section{Introduction}

Topological string theory has proven to be a powerful tool in our understanding of both the physical and mathematical aspects of string theory geometries, particularly in the context of type II string theories and Calabi--Yau compactifications. Following the seminal paper of Witten~\cite{Witten:1988xj}, topological sectors of string theory have led to deep implications in both mathematics and physics, including surprising geometric correspondences such as mirror symmetry~\cite{Candelas:1990rm,Kontsevich:1994dn}, Gopakumar--Vafa invariants and counting BPS states~\cite{Gopakumar:1998ii,Gopakumar:1998jq}, topological open strings~\cite{Witten:1992fb} and Donaldson--Thomas invariants~\cite{thomas1997gauge, donaldson1998gauge}, and invariants for symplectic manifolds~\cite{Witten:1990hr}.

Many of these works have focused on a worldsheet or sigma-model approach, with topological string theory obtained by coupling a twisted worldsheet superconformal field theory to two-dimensional gravity. In the most well-known set-up, the worldsheet theory has $(2,2)$ supersymmetry, and the two choices of twist lead to the A-model and B-model~\cite{Witten:1991zz}. In the A-model case, correlation functions are sensitive only to the K\"ahler class of the target space $X$ of the theory and count the number of rational curves in $X$. In the B-model, correlation functions are sensitive only to the complex structure of the target space. Classical mirror symmetry at genus zero then states that the counting of rational curves on a Calabi--Yau threefold (computed by the A-model) is equivalent to analysing variations of Hodge structure on its mirror Calabi--Yau (computed by the B-model). This correspondence extends to higher-genus worldsheets too. For example, in the A-model, the counting of rational curves is generalised by Gromov--Witten theory to the counting of higher-genus curves. Historically, the higher-genus B-model was less straightforward to understand, as there was no immediate generalisation of variations of Hodge structure to higher genus. However, a worldsheet theory was eventually described by Bershadsky, Cecotti, Ooguri and Vafa (BCOV) in \cite{Bershadsky:1993cx}.

Rather than working solely on the worldsheet of the string, topological theories can also be analysed from the target space of the string. For example, BCOV, put forward a proposal for the target-space theory of the topological B-model, which has come to be known as Kodaira--Spencer theory (or Kodaira--Spencer gravity). Kodaira--Spencer theory characterises the behaviour of complex manifolds and their cohomology under infinitesimal variations. In the context of the worldsheet topological B-model with a Calabi--Yau target space $X$, this is a gauge and gravitational anomaly-free theory which captures crucial aspects of the complex structure moduli space of the string. For example, the periods of the holomorphic $(n,0)$-form on the Calabi--Yau provide a natural set of coordinates on the complex structure moduli space, with infinitesimal changes in these coordinates (variations of Hodge structure) governed by Kodaira--Spencer theory. Put this way, it is clear this target-space theory is best thought of as quasi-topological, since it depends on the complex structure of the target space but not the symplectic structure, and so does not depend on a choice of metric on $X$. Furthermore, classically the theory depends on only holomorphic data of the complex structure, though this holomorphicity does not hold after quantum corrections. 
Continuing this target-space point of view, in the context of the type II string, Pestun and Witten~\cite{Pestun:2005rp} showed that the one-loop partition function of the topological A- and B-models, which agree in the large-volume limit~\cite{Cecotti:1992vy, Bershadsky:1993cx, Bershadsky:1993ta}, can be computed from a quadratic target-space action derived from the Hitchin functional for a generalised Calabi--Yau structure~\cite{Hitchin_2003}.

In this paper, we analyse a similar target-space theory in the context of the heterotic string. Our set-up is compactifications to four-dimensional Minkowski space preserving minimal supersymmetry. The resulting four-dimensional theories are governed by a superpotential~\cite{LopesCardoso:2003dvb, Gurrieri:2004dt,delaOssa:2015maa,McOrist:2016cfl} and a Kähler potential \cite{Anguelova:2010ed, Candelas:2016usb, Garcia-Fernandez:2018emx, Candelas:2018lib, McOrist:2019mxh, Ashmore:2019rkx}. Crucially, the superpotential is a holomorphic functional of the geometry of the compact six-dimensional internal space. Expanding in fluctuations around a given supersymmetric background, the superpotential gives a holomorphic field theory for the moduli of the background.\footnote{While the critical points of the superpotential directly correspond to the F-term supersymmetry conditions, the quotient of this space by complexified gauge equivalence in fact equals the moduli space including the D-terms conditions. See \cite{Ashmore:2018ybe,Ashmore:2019rkx, McOrist:2021dnd} for how this works.} We shall refer to this theory as heterotic Kodaira--Spencer and study its classical and quantum properties. At the classical level, we find that the action can be made independent of the background metric and gauge connection. In particular, it depends only on the holomorphic structure of the manifold and gauge bundle, and the fluxes. We find that our action can be written naturally in the language of \cite{delaOssa:2014cia,Garcia-Fernandez:2015hja,Ashmore:2019rkx}, in which case the theory depends only on the holomorphic structure of the Courant algebroid $(Q,\bar{D})$, and not on the associated metric.
By expanding to quadratic order in fluctuations, we calculate the one-loop contribution to the partition function for this holomorphic field theory. Focusing on the case where the background geometry is Calabi--Yau, we compute this using both traditional operator methods and Batalin--Vilkovisky (BV) quantisation, finding agreement between them, with the answer given in terms of Ray--Singer torsions associated to $(Q,\bar{D})$.

One might wonder about whether this theory is well-defined at the quantum level as, since the theory is (classically) holomorphic, it resembles a chiral theory in six dimensions and so may be anomalous. In particular, the theory may suffer from local anomalies: the classical action governing the fluctuations is invariant under diffeomorphisms and gauge transformations, but the partition function may transform with an anomalous phase. If this cannot be cancelled by local counter-terms, it indicates a fatal breakdown of gauge invariance and means the quantum theory is ill-defined. There is also a second kind of anomaly, similar to the holomorphic anomaly in Kodaira--Spencer theory: the classical theory is metric-independent, but quantisation requires a choice of background metric. The partition function of the theory may then depend on this background metric. If this dependence cannot be cancelled by local counter-terms, as happens for the topological anomaly in Chern--Simons theory,\footnote{As we comment on later, the topological anomaly in Chern--Simons theory can be removed almost entirely, leaving dependence only on a choice of ``framing''.} it indicates that the theory depends on extra data, not present in the classical theory. This is undesirable if one is hoping for a quasi-topological theory, but not fatal. Indeed, a dependence of the one-loop partition function on the background structure might even be desirable, as it may shed light on higher loop corrections through e.g. holomorphic anomaly equations \cite{Bershadsky:1993cx, Bershadsky:1993ta}.

We analyse both kinds of anomalies and study when they can be cancelled by the addition of (local) counter-terms and/or by altering the theory using a Green--Schwarz--like mechanism. We find that the only local anomalies which can occur are gravitational anomalies. We show that these can always be cancelled through a Green--Schwarz mechanism. Interestingly, if we relax some of the conditions one puts into the construction of our theory, we obtain a theory which has gauge, gravitational and mixed anomalies. The cancellation of these requires extra constraints on the topology of the gauge bundles, which we study this in section \ref{sec:untwisted_model}. We also find a non-trivial dependence of the partition function on the background metric and gauge connection. Naively, this breaks the topological nature of the classical theory. However, we find that it can be restored through the addition of counter terms, provided one relates the second Chern characters of the manifold and gauge bundle. We comment on how this relates to the usual constraints on the Chern characters of heterotic backgrounds coming from the Bianchi identity. In particular, the anomaly cancellation relies on the anomaly polynomial factorising in nice ways. To achieve this factorisation it is crucial to include extra fields in the action which can be decoupled from the on-shell moduli problem \cite{Ashmore:2018ybe}. These fields include the axio-dilation and a component of the Kalb-Ramond $B$-field. Understanding how these fields correct the $L_{\infty}$-structure of the moduli problem when included is left for future work.

As our theory is an example of a holomorphic field theory (in the case of vanishing fluxes), we go on to show how our approach to its quantisation can be applied to a more general class of theories of this type, and is closer in spirit to the approach of \cite{thomas1997gauge,Bittleston:2022nfr}.
The field content of a holomorphic field theory is specified by a complex of spaces of holomorphic sections of holomorphic bundles, with holomorphic differential operators providing the differentials~\cite{Williams:2018ows}. 
The BV complex is then given as the total complex of the Dolbeault resolution of this. 
The one-loop partition function is thus expressed in terms of the analytic torsion of this total complex.  
Under mild assumptions, which are fulfilled by a large class of holomorphic field theories on compact K\"ahler manifolds, we show that this analytic torsion in fact decomposes into an alternating product of the holomorphic torsions of the holomorphic bundles in the original complex being resolved. 
Thus, in these cases, the one-loop partition function can be read-off from the field content alone and does not depend on the form of the holomorphic differential appearing in the kinetic terms.  
Further, it will automatically be given in terms of holomorphic torsions, so that one could then perform the anomaly analysis using the formulae for the variations of these holomorphic torsions as we have done for our heterotic theory.

The paper is organised as follows. We begin in Section \ref{sec:het_KS} by introducing the relevant holomorphic field theory, derived from the heterotic superpotential, and then give a computation of the one-loop partition function of this theory using formal path-integral methods and BV quantisation in the cases of both vanishing and non-vanishing flux. In Section \ref{sec:anomaly_review}, we give an overview of both local and ``topological'' anomalies focusing on the examples of Chern--Simons theory and the B-model. In Section \ref{sec:gauge_section}, we discuss local anomalies for the holomorphic field theory governing the moduli of a supersymmetric heterotic background. Analysing the cases of vanishing and non-vanishing flux, we find that the theory is generically anomalous, but the anomalies can be cancelled by a Green--Schwarz mechanism. As an aside, we also discuss a slightly different holomorphic field theory which does not come from the heterotic string. In Sections \ref{sec:metric_anomaly}, we check whether our theory is quasi-topological by analysing the dependence of the one-loop partition function on the background metric and gauge bundle. For particular choices of gauge sector, we find that independence of the background metric and gauge connection can be restored by the addition of local counter-terms. In Section \ref{sec:Torsion}, we discuss how one can understand the previous sections in terms of general properties of holomorphic field theories, finding that the one-loop partition function can be expressed in terms of the Ray--Singer torsions of certain holomorphic bundles. We finish in Section \ref{sec:conclusions} with some discussion of future directions.

\section{Heterotic Kodaira--Spencer theory}\label{sec:het_KS}

We begin by introducing the theory in question, derived from the heterotic superpotential. As we will see, this theory governs supersymmetric deformations of heterotic string theory on a complex threefold, and is closely related to Kodaira--Spencer gravity \cite{Bershadsky:1993cx}, and holomorphic Chern--Simons theory \cite{thomas1997gauge, donaldson1998gauge}.

\subsection{The superpotential, deformations and Maurer--Cartan}

Our starting point is to consider deformations of the Hull--Strominger system, which describes $\cN = 1$ supersymmetric reductions of heterotic theory on a six-dimensional manifold with an $\SU{3}$ structure. A heterotic Kodaira--Spencer theory should, at the classical level, reproduce the Maurer--Cartan equations of on-shell, or ``integrable'', deformations and hence can be considered the heterotic analogue of Kodaira--Spencer theory for deformations of complex structures \cite{Bershadsky:1993cx}.\footnote{Here we use integrable in the sense of \cite{Ashmore:2019rkx} to mean deformations preserving supersymmetry. This is somewhat non-standard compared with the literature on $G$-structures.}  Such a theory was written down in \cite{Ashmore:2018ybe}, where it was found to be described by an $L_{\infty}$-algebra associated to the underlying gauge structure. We will review this work and extend it to allow for deformations which include rescalings of the holomorphic three-form $\Omega$.

Supersymmetric compactifications of heterotic string theory on a six-dimensional manifold $X$ require that the internal geometry admits an $\SU{3}$ structure~\cite{Hull:1986kz,Strominger:1986uh}. The $\SU{3}$-invariant tensors $(\omega,\Omega)$ and the gauge field $A$ must satisfy certain differential conditions for the background to preserve supersymmetry:
\begin{gather}
        \dd\Omega = 0 \, , \qquad \dd^{J} \omega = H \coloneqq \dd B + \tfrac{1}{4}\alpha'\left(\omega_{\text{CS}}(A) - \omega_{\text{CS}}(\Theta)\right) \, , \qquad F_{0,2} = R_{0,2} = 0\, ,\label{eq:HS_F_terms}\\
        \dd(\ee^{-2\varphi}\omega \wedge \omega) = 0 \, , \qquad \omega\lrcorner F = \omega^{}\lrcorner R = 0 \, ,\label{eq:HS_D_terms}
\end{gather}
where $J$ is the complex structure associated to $(\omega,\Omega)$.
These equations imply that the background is complex with a hermitian metric that is generically not K\"ahler, but only conformally balanced. Both the tangent bundle and gauge bundle must be holomorphic bundles equipped with hermitian Yang--Mills connections. The three-form flux $H$ is defined in terms of the Chern--Simons three-form associated to the gauge connection $A$ and the spin connection $\Theta$. This ensures that $H$ satisfies the usual non-trivial Bianchi identity for heterotic backgrounds. Due to the difficulty of working with the spin connection explicitly,\footnote{See e.g. \cite{McOrist:2021dnd}.} we will ignore it and consider only the contribution of the gauge field $A$ to $H$. This is a toy model for heterotic backgrounds which is frequently employed \cite{Ashmore:2019rkx,Ashmore:2018ybe,Garcia-Fernandez:2018ypt,Garcia-Fernandez:2018emx}. One could assume that the gauge group has an $\SU{3}$ factor and impose by hand at the end of the calculation that the associated gauge connection aligns with the spin connection of the manifold.\footnote{One also should take care about the relative sign between the gauge sector and the spin connection in the definition of $H$} However, one needs to be careful about spurious degrees of freedom arising from treating the gauge field as independent from the hermitian metric, particularly when doing 1-loop calculations. We shall ignore this subtlety here.

The four-dimensional $\mathcal{N}=1$ effective theory one obtains after compactifying on such a background is controlled by a K\"ahler potential and a superpotential. If one keeps full dependence on the internal geometry (rather than truncating to a finite set of modes), these $\mathcal{N}=1$ potentials are functionals of the internal geometry. Explicitly, the superpotential functional $W$ is given by~\cite{LopesCardoso:2003dvb, Gurrieri:2004dt}
\begin{equation}
\label{eq:superpotential}
W=\int_X\left(H+\ii\,\dd\omega\right)\wedge\Omega \, .
\end{equation}
This functional is properly identified as ``the'' superpotential by noting that $W=\delta W=0$ is equivalent to the F-term conditions of the Hull--Strominger system~\cite{LopesCardoso:2003dvb, delaOssa:2015maa, McOrist:2016cfl}. These are the conditions appearing in \eqref{eq:HS_F_terms}, while those appearing in \eqref{eq:HS_D_terms} are the D-terms. We will often refer to field configurations that satisfy the F-term conditions, $W=\delta W=0$, as being ``on-shell''. Though one also needs to solve the D-term conditions to have a solution to the equations of motion, the space of solutions to the F-term equations modulo complexified gauge transformations (which are included in our construction below) provides the physical moduli space~\cite{Ashmore:2018ybe,Ashmore:2019rkx}, so that our ``on-shell" states are in one-to-one correspondence to the truly supersymmetric states.

In analogy to Kodaira--Spencer theory, we now look for an action whose equations of motion describe on-shell deformations of the background. To do so, we expand \eqref{eq:superpotential} for a generic deformation of the background degrees of freedom $(\Omega, \omega,B,A)$. For now, we will ignore the gauge fields\footnote{Or equivalently, work in the large-volume limit where $\alpha'$ corrections are suppressed.} and consider deformations of the geometry alone. We can parametrise deformations of the complex structure by a complex scalar $k$ and a Beltrami differential $\mu \in \Omega^{0,1}(T^{1,0})$. Here our notation is that $\Omega^{p,q}(V)$ is the space of $V$-valued $(p,q)$-forms on $X$. Combining the deformations of $\omega$ and $B$ into complex objects, one finds finite deformations of $(\Omega,\omega,B)$ can be expressed as
\begin{equation}\label{eq:finite_defs}
    \begin{aligned}
        \Omega + \Delta\Omega &= (1+k)(\Omega + \imath_{\mu} \Omega + \tfrac{1}{2}\imath_{\mu}\imath_{\mu}\Omega + \tfrac{1}{3!}\imath_{\mu}\imath_{\mu}\imath_{\mu}\Omega ) \, ,\\
        (\Delta B + \ii \Delta \omega)_{1,1} &= x \, ,\\
        (\Delta B + \ii \Delta \omega)_{0,2} &= b\, .
        \end{aligned}
\end{equation}
Here, $\imath_{\mu}$ denotes contraction with the vector index of $\mu$ and antisymmetrising with the form index as usual. Note that it was shown in \cite{Candelas:2016usb, Ashmore:2018ybe} that one can always set $(\Delta B + \ii\Delta \omega)_{2,0} = 0$. 

The heterotic Kodaira--Spencer theory we consider is then given by
\begin{equation}\label{eq:het_KS_action}
    \begin{aligned}
        S \equiv W + \Delta W
        &= \int_{X} \tfrac{1}{2}(H+\ii\,\dd\omega)\wedge \imath_{\mu}\imath_{\mu}\Omega + (\delb x+\del b )\wedge \imath_{\mu}\Omega + \delb b \wedge k\Omega \\
        & \eqspace +\int_{X} \tfrac{1}{2}(H+\ii\,\dd\omega)\wedge k\imath_{\mu}\imath_{\mu} \Omega + (\delb x + \del b) \wedge k \imath_{\mu}\Omega + \tfrac{1}{2} \del x \wedge \imath_{\mu}\imath_{\mu} \Omega \\
        & \eqspace + \int_{X}\tfrac{1}{2} \del x \wedge k\imath_{\mu}\imath_{\mu}\Omega \, ,
    \end{aligned}
\end{equation}
where in the second equality we have used the fact that we are deforming around an on-shell background to remove the terms at zeroth- and first-order in the deformations. The first, second and third lines correspond to the terms in the action which are quadratic, cubic and quartic in the deformations. As shown in \cite{delaOssa:2015maa,Ashmore:2018ybe}, the resulting equations of motion of this action give the moduli equations for the Hull--Strominger system.

We can package the deformations in a way that reflects the underlying $L_{\infty}$ structure of the moduli problem. We first define the spaces $A^p$ as
\begin{equation}\label{eq:Q_def}
    A^{p} = \Omega^{0,p}(Q) \oplus \Omega^{0,p+1} \oplus \Omega^{0,p-1} \, , \qquad Q = T^{1,0}\oplus  T^{*1,0}\, .
\end{equation}
for $p=0$, the last summand is not present, while for $p=-1$, only the middle summand appears. Writing $y=(\mu,x,b,k)$, we see that elements $y\in A^{1}$ capture deformations of the background, as in \eqref{eq:finite_defs}. There is then a natural (holomorphic) pairing $\langle \cdot,\cdot \rangle $ and a differential $\mathcal{D}$ on the complex $\mathcal{A} \equiv \bigoplus_{p}A^{p}$:
\begin{equation}
        \langle \cdot,\cdot \rangle  \colon A^{p}\times A^{3-p} \to  \Omega^{0,3} \, , \qquad \mathcal{D}\colon A^{p} \to A^{p+1}\, .
\end{equation}
Explicitly, for $y \in A^{p}$, $y'\in A^{3-p}$, these are given by
\begin{equation}
    \begin{aligned}
        \langle y,y'\rangle &= -\imath_{\mu} x' + \imath_{\mu'}x + k\wedge b' - k' \wedge b \, ,\\
        \mathcal{D}y &= \bigl(\delb \mu, \delb x + \del b +(-1)^p\,\imath_{\mu}\tilde{H}, \delb b, \delb k + (-1)^{p} \del \cdot \mu\bigr)\, ,
    \end{aligned}
\end{equation}
where $\tilde{H} \equiv H+\ii\,\dd\omega \in \Omega^{2,1}$, and $\del\cdot \mu$ denotes the holomorphic divergence defined using the holomorphic volume form $\Omega$ on $X$. It is then straightforward to check that $\mathcal{D}$ is nilpotent and (graded) self-adjoint with respect to the pairing:
\begin{equation}
    \mathcal{D}^{2} = 0 \, , \qquad \int_{X}\langle \mathcal{D}y, y' \rangle  \wedge \Omega = (-1)^{p}\int_{X} \langle y,\mathcal{D}y'\rangle \wedge \Omega\, ,
\end{equation}
where $y\in A^p$.

Using this repacking of the deformations, the quadratic part of the action \eqref{eq:het_KS_action} can be written as
\begin{equation}\label{eq:quadratic_action}
    S_{2} = \frac{1}{2}\int_{X} \langle y,\mathcal{D} y\rangle  \wedge \Omega \, , \qquad y \in A^{1}\, .
\end{equation}
It is clear from this expression that this theory is independent of a choice of background metric. It only depends on a choice of $\tilde{H}$ which defines the extension structure of $Q$. In principle, one can find higher order brackets $l_{2},l_{3},...$ such that $\mathcal{A}$ admits an $L_{\infty}$ structure, with the full action then written in terms of $(\mathcal{D},l_{2},l_{3},\dots)$. In this paper, however, we will be interested only in the one-loop partition function of the theory controlling the deformations. This requires only the quadratic action, so we will leave the derivation of the full $L_{\infty}$ algebra for future work.\footnote{As mentioned already, the $L_{\infty}$ algebra for deformations not including rescalings of $\Omega$ was derived in \cite{Ashmore:2018ybe}.} Finally, we note that the quadratic action \eqref{eq:quadratic_action} has a gauge symmetry given by
\begin{equation}\label{eq:gauge_sym_i}
    \delta y = \mathcal{D} y_{0} \, , \qquad y_{0} = (\mu_{0},x_{0},b_{0},0) \in A^{0}\, ,
\end{equation}
and a gauge-for-gauge symmetry
\begin{equation}
    \delta y_{0} = \mathcal{D} y_{-1} \, , \qquad y_{-1} = (0,0,b_{-1}, 0) \in A^{-1}\, .
\end{equation}

\subsection{The one-loop partition function}\label{sec:one-loop}

Given the heterotic Kodaira--Spencer action \eqref{eq:het_KS_action}, we would like to understand its quantisation. The procedure we will follow is similar to what one does in conventional Chern--Simons theory on a three-manifold. There, instead of trying to compute the path integral by gauge fixing the action globally, one fixes a critical point of the Chern--Simons functional and then gauge fixes the theory around it. In a little more detail, critical points correspond to flat connections, $\dd A_0 + A_0\wedge A_0 = 0$. One then writes the Chern--Simons field $A$ as a fluctuation $a$ around such a flat connection, $A=A_0 + a$. The action can then be written as
\begin{equation}
    S_{\text{CS}}[A] = S_{\text{CS}}[A_0] + \int\tr\left(a\wedge \dd_{A_0}a + \tfrac{2}{3}a\wedge a \wedge a \right)\equiv S_{\text{CS}}[A_0] + S^{A_0}_{\text{CS}}[a]\, .
\end{equation}
The partition function of this theory factorises as~\cite{hep-th/9110056}
\begin{equation}\label{eq:cs_partition_function}
    Z_{\text{CS}} =\int \dd A_{0}\, Z_{\text{back}} \, Z_{\text{free}} \, Z_{\text{pert}},
\end{equation}
where $Z_{\text{back}}=\exp(-S_{\text{CS}}[A_0])$, $Z_{\text{free}}$ is the partition function for the quadratic part of $S^{A_0}_{\text{CS}}[a]$, and $Z_{\text{pert}}$ is the contribution from treating the cubic term in $S^{A_0}_{\text{CS}}[a]$ perturbatively. We have also included an explicit integral over the zero-modes of the field. The terms $Z_{\text{back}}$ and $Z_{\text{free}}$ are usually called the zero- and one-loop parts of the partition function respectively. To properly quantise the action, one needs to gauge fix which requires the introduction of a background metric $g$ on the underlying three-manifold. The topological nature of the classical theory is then naively broken in the quantum theory, though this can be restored up to a choice of framing~\cite{Witten:1988hf}. We will discuss this in a little more detail in the following sections when we consider anomalies. At the free level, the absolute value of the partition function can be expressed in terms of the Ray--Singer torsion \cite{ray1971r}, while the phase is determined by the (twisted) $\eta$-invariant \cite{Atiyah:1975jf,Atiyah:1976jg,Atiyah:1976qjr} of the three-manifold and gauge bundle for $A$.

\subsubsection*{Heterotic Kodaira--Spencer}

For us, the analogue of $A_{0}$ is a choice of on-shell background geometry $(\Omega,\omega,B)$. Since the zeroth-order action is the superpotential functional, and this vanishes for on-shell backgrounds, the background contribution to the partition function is trivial, $Z_{\text{back}}=1$. The one-loop partition function $Z_{\text{free}}$ is the contribution from the quadratic action \eqref{eq:quadratic_action}. That is, it is the partition function of the free theory viewed as a perturbation around some fixed on-shell background. When evaluated, one expects it will give a geometric invariant of the background geometry $(\Omega,\omega,B)$. In fact, as we will see in Section \ref{sec:metric_anomaly}, any dependence on the hermitian structure can be removed through appropriate counter-terms and hence $Z_{\text{free}}$ calculates complex invariants of the background. 

For the rest of this work, we will not consider the interaction terms which lead to $Z_{\text{pert}}$. We will also ignore the contribution from $\mathcal{D}$-harmonic modes. These zero-modes correspond to classical moduli, i.e.~changes in the background geometry, and give $Z_{\text{free}}$ the interpretation of a measure on $\mathcal{M}_{\text{het}}$. Note that the analogue of \eqref{eq:cs_partition_function} for heterotic Kodaira--Spencer will include an integral of $Z_{\text{free}}$ over the zero-modes, i.e.~the moduli space $\mathcal{M}_{\text{het}}$. This should promote the complex invariants $Z_{\text{free}}$ to topological invariants of $X$. We do not attempt this and instead focus on calculating $Z_{\text{free}}$ without zero-modes.\footnote{We cannot completely neglect the zero-modes as these give rise to potential gauge anomalies. We will discuss this further in the next section.} As stated in the introduction, since our action is complex, the partition function as stated is ill-defined. We shall discuss this issue and its resolution at the end of this section.

\subsection{Computing \texorpdfstring{$Z_{\text{free}}$}{Z} for \texorpdfstring{$H=0$}{H = 0}}\label{sec:H=0}

Let us, for now, set $H=0$ which implies that $X$ is Calabi--Yau. In this case, it is easier to redefine our variables to absorb the factor of $\Omega$. In particular, we will define $\chi = \imath_{\mu}\Omega$, $\kappa = k\,\Omega$. We simultaneously redefine $\mathcal{A}$, $\mathcal{D}$ and the pairing $\langle \cdot,\cdot\rangle $ such that
\begin{equation}\label{eq:L_inft_redef}
    \begin{aligned}
        A^{p} &= \Omega^{2,p}\oplus \Omega^{1,p} \oplus \Omega^{0,p+1} \oplus \Omega^{3,p-1} \, ,\\
        \mathcal{D}y &= (\delb \chi,\, \delb x + \del b, \,\delb b ,\, \delb \kappa + \del \chi ) \, ,\\
        \langle  y_, y' \rangle  &= x\wedge \chi' - x'\wedge \chi + b \wedge \kappa' - b'\wedge \kappa \, ,
    \end{aligned}
\end{equation}
where $y\in A^{p}$, $y'\in A^{3-p}$. We are then looking to quantise the action
\begin{equation}\label{eq:quadratic_action_ii}
    S_{2} = \tfrac{1}{2} \int_{X} \langle y,\mathcal{D}y\rangle  = \int_{X} \left((\delb x + \del b)\wedge \chi + \delb b\wedge \kappa \right)\, .
\end{equation}

Before performing full BV quantisation, we give a formal calculation of the partition function, which is useful for intuition. Following the approach of Schwarz~\cite{Schwarz:1979ae,Schwarz:1978cn}, formally, we can write
\begin{equation}\label{eq:Z_free_formal}
    Z_{\text{free}} = \frac{1}{\Vol(G)} \int Dy\, \ee^{-S_{2}[y]} = \frac{1}{\Vol(G)}\, \frac{\Vol (\mathcal{D}A^{0})}{ \det(\mathcal{D}|_{A^{1}})^{1/2}}\, ,
\end{equation}
where $\Vol(G)$ is the formal volume of the gauge group. We then note the identity
\begin{equation}
    \Vol (\mathcal{D} A^{0}) = \frac{ \det(\mathcal{D}|_{A^{0}})  \Vol(A^{0})}{ \Vol(\mathcal{D}A^{-1}) } = \frac{ \det(\mathcal{D}|_{A^{0}})  \Vol(A^{0})}{ \det(\mathcal{D}|_{A^{-1}})  \Vol(A^{-1}) } \, ,
\end{equation}
and hence, provided we define $\Vol(G) = \Vol(A^{0})/\Vol(A^{-1})$, the one-loop partition function is simply
\begin{equation}\label{eq:Z_free_D}
    Z_{\text{free}} = \frac{ \det(\mathcal{D}|_{A^{0}}) }{ \det(\mathcal{D}|_{A^{-1}})  \det(\mathcal{D}|_{A^{1}})^{1/2} }\, .
\end{equation}

Since $\mathcal{D}$ is not an endomorphism on $A^{p}$, a definition of $\det(\mathcal{D}|_{A^{p}})$ is given by\footnote{See, for instance, the discussion in \cite{Pestun-Witten2005}.}
\begin{equation}\label{eq:DDdagger}
    \det(\mathcal{D}|_{A^{p}}) = \ee^{\ii\phi(\mathcal{D})}  \det(\mathcal{D}^{\dagger}\mathcal{D}|_{A^{p}})^{1/2} \, .
\end{equation} 
for some phase $\phi(\mathcal{D})$ which we will return to later. Thus, we need to define the adjoint operator $\mathcal{D}^{\dagger}$, which in turn requires an inner product $(\cdot,\cdot)$ on $\mathcal{A}$. We do so by introducing a conventional metric $g$ on $X$ which we take to coincide with the K\"ahler metric of the background.\footnote{This seems like the natural choice, although we note that it need not be the case. For example, one could consider using a more general metric on the bundle $Q=T^{1,0}\oplus  T^{*1,0}$.} The inner product on $y,y'\in A^{p}$ is then given by
\begin{equation}
\label{eq:het+ve-prod}
    \begin{aligned}
        (y,y') &= \langle  y,\overline{* y'} \rangle  = (\chi,\chi') + (x,x') + (b,b') + (\kappa,\kappa')\, ,
    \end{aligned}
\end{equation}
where, on the right-hand side, we have used $(\cdot,\cdot)$ to denote the usual inner product on differential forms. One can then calculate the adjoint operator $\mathcal{D}^{\dagger}$ via
\begin{equation}
    \begin{aligned}
        (\mathcal{D} y, y') &= (\delb \chi, \chi') + (\delb x + \del b, x') + (\delb b, b') + (\delb \kappa + \del \chi, \kappa') \\
        &= (\chi, \delb^{\dagger}\chi'+ \del^{\dagger}\kappa') + (x, \delb^{\dagger} x') + (b, \delb^{\dagger} b' + \del^{\dagger} x') + (\kappa, \delb^{\dagger} \kappa') \\
        &= (y, \mathcal{D}^{\dagger} y')\, ,
    \end{aligned}
\end{equation}
where now $y\in A^p$ and $y'\in A^{p+1}$, which results in 
\begin{equation}\label{eq:D_dagger}
    \mathcal{D}^{\dagger}y = \left(\delb^{\dagger}\chi + \del^{\dagger} \kappa,\, \delb^{\dagger}x,\, \delb^{\dagger}b + \del^{\dagger}x,\, \delb^{\dagger}\kappa\right)\, .
\end{equation}

With this, we can evaluate \eqref{eq:Z_free_D} in terms of determinants of conventional Laplacians.\footnote{Here and throughout the paper, we will work with $\zeta$-regularised determinants. In addition, all determinants are assumed to be computed with any zero-modes removed.} To do so, we rely on the fact that the inner product on $\mathcal{A}$ defines a Hodge decomposition
\begin{equation}
    A^{p} = \mathcal{H}^{p}\oplus \mathcal{D}A^{p-1} \oplus \mathcal{D}^{\dagger}A^{p+1}\, ,
\end{equation}
where $\mathcal{H}$ is the space of $\mathcal{D}$-harmonic modes, which we ignore for this calculation. The Laplacian $\mathcal{L} = \mathcal{D}\mathcal{D}^{\dagger} + \mathcal{D}^{\dagger}\mathcal{D}$ acting on $A^{p}$ then splits into $\mathcal{D}\mathcal{D}^{\dagger}$ acting on $\mathcal{D}A^{p-1}$ and $\mathcal{D}^{\dagger}\mathcal{D}$ acting on $\mathcal{D}^{\dagger}A^{p+1}$. Hence, the determinant of the Laplacian can be formally expressed as
\begin{equation}
    \det(\mathcal{L}) = \det(\mathcal{D}\mathcal{D}^{\dagger})\det(\mathcal{D}^{\dagger}\mathcal{D})\, .
\end{equation}
A further calculation also shows that
\begin{equation}
    \det(\mathcal{D}\mathcal{D}^{\dagger}|_{A^{p}}) = \det (\mathcal{D}^{\dagger}\mathcal{D}|_{A^{p-1}})\, .
\end{equation}
From this, we can rewrite the one-loop partition function \eqref{eq:Z_free_D} in terms of determinants of Laplacians as
\begin{equation}\label{eq:Z_free_L}
    \left|Z_{\text{free}}\right| = \frac{ \det(\mathcal{L}|_{A^{0}})^{3/4}}{ \det(\mathcal{L}|_{A^{-1}})^{5/4} \det(\mathcal{L}|_{A^{1}})^{1/4}}\, .
\end{equation}

Finally, we can use the definition of $\mathcal{D}^\dagger$ from \eqref{eq:D_dagger} and the equality of the Dolbeault Laplacians on a K\"ahler manifold, $\Delta_{\delb} = \Delta_{\del} = \tfrac{1}{2}\Delta$, to simplify the $\mathcal{D}$-Laplacian as
\begin{equation}
    \mathcal{L} = 2\del^{\dagger}\del + \del\del^{\dagger} \qquad \Rightarrow \qquad \det(\mathcal{L}) = \det(2\del^{\dagger}\del)\det(\del\del^{\dagger})\, .
\end{equation}
Rescaling an operator by a constant just multiplies the determinant by a constant.\footnote{In the $\zeta$-function regularisation scheme we have chosen, provided rescaling does not change the domain of an operator (which is true in the cases we consider), for any constant $a$ we have
\begin{equation*}
    \det(a\,\mathcal{O}) = a^{\zeta(0)}\det \mathcal{O}\, .
\end{equation*}} Hence, up to some constant factor which can be absorbed into the definition of $Z_\text{free}$, we have
\begin{equation}
\label{eq:det-equality}
    \det(\mathcal{L}) = \det(\Delta)\, .
\end{equation}
Plugging this into \eqref{eq:Z_free_L}, decomposing the bundles $A^{p}$ according to \eqref{eq:L_inft_redef}, and using the results in Appendix \ref{app:Hodge}, one finds%
\begin{equation}\label{eq:Z_free_Delta}
    \left|Z_\text{free}\right| = \frac{ \det(\Delta_{1,0})^{2}}{ \det(\Delta_{0,0})^{3/2}\det(\Delta_{1,1})^{1/2} } = \frac{B}{C^{1/2}} \, .
\end{equation}
In the first equality, we have used $\Delta_{p,q}$ to denote the conventional de Rham Laplacian restricted to $\Omega^{p,q}$, and in the second equality, we have introduced the quantities $B$ and $C$, defined in Appendix \ref{app:Hodge}, which are the restriction of $\det(\Delta)$ to certain subspaces within the Hodge diamond of $X$. (See Figure \ref{fig: Hodge diamond} for details.)

\subsubsection*{Ray--Singer torsion and the partition function}

One may worry at this point whether the path-integral expression in \eqref{eq:Z_free_formal} is well-defined as, naively, any path integral with a complex action appears to be divergent. To highlight this issue, we take as a toy example the one-dimensional integral
\begin{equation}\label{eq:complex_gaussian_integral}
    I = \frac{1}{\sqrt{\pi}}\int_{\mathcal{C}} \ee^{-\lambda z^{2}}\dd z\, ,
\end{equation}
with $\lambda \in \mathbb{R}_{>0}$. This integral is defined only once we prescribe a contour $\mathcal{C}$. Taking $\mathcal{C}=\mathcal{C}_{0}$ to be the real line, we find the familiar result $I_{0}=\lambda^{-1/2}$. If, however, we rotate the contour $\mathcal{C}=\mathcal{C}_{\phi}$ to have a phase $\ee^{\ii\phi}$ for $\phi \in (-\tfrac{\pi}{4},\tfrac{\pi}{4})$, we find that $I$ also picks up a phase:
\begin{equation}\label{eq:phi_contour}
    I_{\phi}=\ee^{-\ii\phi}\lambda^{-1/2}\, .
\end{equation}
If we take $\phi$ to be outside of this range, the integral may no longer converge. However, even in this case, we can use \eqref{eq:phi_contour} as a \emph{definition} of the value of the integral for all $\phi$.\footnote{See, for example, \cite[Appendix A]{thomas1997gauge}.} Note that, with this definition, the absolute value of the integral is independent of the contour and is given by $\lambda^{-1/2}$. We can make a similar definition for the infinite-dimensional integral \eqref{eq:Z_free_formal} which defines $Z_\text{free}$. After decomposing into eigenstates of $\mathcal{D}$, we have an infinite product of integrals of the form \eqref{eq:complex_gaussian_integral}, where now $\lambda$ can be positive or negative. This integral is well-defined only after choosing some real slice of the complex field space over which to integrate. Despite this, the absolute value of the integral, given by the product of determinants of Laplacians as in \eqref{eq:Z_free_Delta}, should be well-defined and independent of the choice of contour.

The particular combination of Laplacians appearing in \eqref{eq:Z_free_Delta} is significant because the absolute value of $Z_{\text{free}}$ can be written as a product of holomorphic Ray--Singer torsions \cite{bismut1988analytic1,bismut1988analytic2,bismut1988analytic3}. Given a holomorphic bundle $V\rightarrow X$ with hermitian metric $h$, one defines the holomorphic Ray--Singer torsion of $V$ to be
\begin{equation}
    I(V) = \prod_{p=0}^{3}\left( \det(\Delta^{V}_{p})^{(-1)^{p+1}p} \right)^{1/2}\, ,
\end{equation}
where $\Delta^{V}_{p}$ is the Dolbeault Laplacian defined with respect to the Chern connection of $h$ on $\Lambda^{0,p}(V)$. With this, on a Calabi--Yau background, the absolute value of the one-loop partition function is
\begin{equation}\label{eq:Z_free_RS_torsions}
    |Z_{\text{free}}| = \frac{I(Q)^{1/2}}{I(\Lambda^{0,0})}\, ,
\end{equation}
where, as in \eqref{eq:Q_def}, $Q=T^{1,0}\oplus  T^{*1,0}$. It is interesting to note that this is the same as the value of the one-loop partition function of the $\SL{3,\mathbb{C}}$ Hitchin functional, as found in \cite{Pestun-Witten2005}. This is indicative of the fact that $Z_{\text{free}}$ calculates holomorphic invariants of the background.

The phase of the partition function is difficult to calculate as it will be highly sensitive to the real slice we choose, as in the one-dimensional example above. For this reason, we will mostly focus on the absolute value of $Z_\text{free}$. However, we note briefly that if we change the real slice by a phase, we pick up an overall phase in the partition function which is proportional to the sum of the signs of the eigenvalues of the operator $\slashed{\mathcal{D}}=*\mathcal{D} + \mathcal{D}*$. After regularising, this sum is simply the $\eta$-invariant for $\slashed{\mathcal{D}}$. That is, under a field redefinition one has
\begin{equation}\label{eq:phase_change}
    y \rightarrow \ee^{\ii\phi}y \qquad \Rightarrow \qquad Z_{\text{free}} \rightarrow \ee^{-\ii\phi(\eta(\slashed{\mathcal{D}}) + \zeta(0))}Z_{\text{free}}\, ,
\end{equation}
where the $\zeta(0)$ term comes from the ghost sector in the BV quantisation. In the case studied where $H=0$, this becomes the $\eta$-invariant of the operator $*\delb+\delb *$, which has nice properties on K\"ahler manifolds. It is interesting that we find a relationship between the phase of the partition function and an $\eta$-invariant. We view the ambiguity in the choice of real slice much like how the frame ambiguity is viewed in Chern--Simons \cite{Witten:1988hf}. Provided we define the partition function and how it changes under a change of phase, we say the partition function is well-defined.

Finally, we note that, much like in Chern--Simons theory, quantisation required the introduction of a metric. Despite the fact that the classical action \eqref{eq:quadratic_action} depends only on the complex structure, the quantum theory naively has metric dependence, and hence we may have an anomaly associated to the topological symmetry. We will see this metric appearing also when we BV quantise the theory. We will analyse the metric dependence of the theory in Section \ref{sec:metric_anomaly}.

\subsubsection*{BV quantisation}

For completeness, we also compute the partition function using the Batalin--Vilkovisky (BV) formalism~\cite{Batalin:1981jr}. To obtain the BV action, we take the quadratic action \eqref{eq:quadratic_action_ii} and construct the corresponding master action. To do this, we introduce the ghosts $(y_0, y_{-1})$ and antighosts $(y_2, y_{3}, y_4)$, where $y_i\in A^i$ with statistics $(-1)^{i+1}$.

The gauge symmetries of the theory fix the action of the BRST operator $\Q$
\begin{equation}
    \Q y_i= \Dc y_{i-1}\, .
\end{equation}
The master action is then given by
\begin{equation}
\begin{aligned}
    S_{\text{BV}}&= S_2+ \int_X \bigl( \langle y_2, \Q y_1 \rangle + \langle y_{3},\Q y_{0} \rangle \bigr)\\
    &= \int_X \bigl(\tfrac12 \langle y_1, \Dc y_1 \rangle + \langle y_2, \Dc y_0 \rangle +\langle y_{3},\Dc y_{-1}\rangle\bigr)\, .
    \end{aligned}
\end{equation}
The next step is to choose a Lagrangian submanifold $L$ in the space of fields. It is convenient to pick this by projecting every field to a subspace orthogonal to its gauge variation:
\begin{equation}
   L = \{y\:|\: \Dc^{\dagger} y =0\}. 
\end{equation}
Assuming that the cohomology associated to $\Dc$ vanishes, the Lagrangian subspace is spanned by $y= \Dc ^{\dagger} u$. Since $\Dc$ has a standard Hodge decomposition, the operator $\Dc$ has no kernel on this subspace. Thanks to this, and noting that $y_1$ and $y_{-1}$ are bosonic and $y_0$ is fermionic, the partition function of the theory is (up to constant factors)
\begin{equation}
    \left|Z_\text{free}\right|=\frac{\det(\Dc|_ {A^0})}{\det(\Dc |_{A^1})^{1/2} \det(\Dc|_ {A^{-1}})}\, .
\end{equation}
This agrees with our earlier calculation of $Z_\text{free}$ in \eqref{eq:Z_free_D}.

\subsection{Including the gauge fields}\label{sec:het_KS_gauge_fields}

We now consider the action when we include the gauge sector of the heterotic string. We will also drop any assumptions on the flux $H$, so in particular, our background may no longer be K\"ahler. Upon turning on the gauge fields, we add a Chern--Simons term $\omega_{\text{CS}}(A)$ to the definition of $H$. Varying the action, one finds an additional term which depends on the gauge fields, given by
\begin{equation}
    S^{\text{gauge}} = \int_{X}\Delta \omega_{\text{CS}}(A)\wedge (1+k)(\Omega+ \imath_{\mu}\Omega + \tfrac{1}{2} \imath_{\mu}\imath_{\mu}\Omega + \tfrac{1}{3!}\imath_{\mu}\imath_{\mu}\imath_{\mu}\Omega)\, .
\end{equation}
Taking $\Delta A = \alpha \in \Omega^{0,1}(\End (V))$, the variation of the Chern--Simons term is simply \cite{Ashmore:2018ybe}
\begin{equation}
    \Delta\omega_{\text{CS}}(A) = 2\tr(F\wedge \alpha) + \dd\tr(\alpha\wedge A) + \tr(\alpha\wedge \dd_{A}\alpha) + \tfrac{2}{3}\tr(\alpha\wedge\alpha\wedge\alpha)\, ,
\end{equation}
and hence the quadratic part of the action is given by
\begin{equation}
    S_{2}^{\text{gauge}} = \int_{X} \tr\left[ k\,\delb(\alpha\wedge A) - \imath_{\mu}\bigl(2F\wedge \alpha + \dd(\alpha\wedge A)\bigr) + \alpha\wedge \delb_{A} \alpha \right]\wedge \Omega\, .
\end{equation}
We can then redefine $x$ and $b$ to absorb the $\dd(\alpha\wedge A)$ term via
\begin{equation}
    x \to x + \tr(\alpha\wedge A_{1,0}) \, , \qquad b \to b+ \tr(\alpha\wedge A_{0,1})\, .
\end{equation}
The result of this is that the sum of the action in Equation \eqref{eq:quadratic_action} and the new contribution from the gauge fields can be written as
\begin{equation}\label{eq:quadratic_action_full}
\begin{aligned}
    S_{2} + S_{2}^{\text{gauge}} &= \int_{X} \left[\imath_{\mu}\bigl(\tfrac{1}{2}\imath_{\mu}\tilde{H} - (\delb x+\del b ) - 2\tr(F\wedge \alpha) \bigr)    + k\,\delb b +\tr( \alpha\wedge \delb_{A} \alpha) \right]\wedge\Omega \\
    &\equiv \tfrac{1}{2} \int_{X} \langle y,\mathcal{D} y \rangle \wedge \Omega\, ,
\end{aligned}    
\end{equation}
where in the second equality, we have are using notation similar to \eqref{eq:quadratic_action}. From the above expression, we can see that if we choose the Chern connection, then the theory does not depend explicitly on $A$. It depends only on the curvature $F$ which, along with $\tilde{H}$, defines the extension structure of the bundle $Q$ below. We can again build a BRST complex $A^{p}$ as in the previous section, where now the $A^{p}$ pick up terms depending on the gauge bundle
\begin{equation}
    A^{p} = \Omega^{0,p}(Q)\oplus \Omega^{0,p+1}\oplus \Omega^{0,p-1} \, , \qquad Q = T^{1,0}\oplus \End(V) \oplus  T^{*1,0}\, .
\end{equation}
As in \eqref{eq:gauge_sym_i}, the physical fields are $y \in A^{1}$, while the gauge parameters are $y_0\in A^{0}$ with the gauge transformations generated by $\delta y = \mathcal{D} y_0$. The pairing $\langle \cdot,\cdot \rangle $ and differential $\mathcal{D}$ also pick up extra terms from the gauge sector. For $y \in A^{p}$ and $y'\in A^{3-p}$ we get the inner product
\begin{equation}
\label{eq:gauge_ip_def}
        \langle y,y' \rangle  = -\imath_{\mu} x' + \imath_{\mu'} x + k\wedge b' -k'\wedge b + 2\tr(\alpha\wedge \alpha') \, ,
\end{equation}
and the differential is given by
\begin{equation}
\label{eq:gauge_D_def}
        \resizebox{.91\hsize}{!}{$\mathcal{D} y = \bigl( \delb \mu, \, \delb x + \del b - 2(-1)^p\tr(F\wedge \alpha) +(-1)^p \imath_{\mu} \tilde{H}, \, \delb_{A} \alpha +(-1)^p  \imath_{\mu}F, \, \delb b, \,\delb k + (-1)^{p}\del\cdot \mu\bigr)\, .$}
\end{equation}
Note that, upon restricting to $(\mu,x,\alpha,b)$, we recover the differential $\bar{D}$ of \cite{Gualtieri:2010fd, delaOssa:2014cia, Anderson:2014xha, Ashmore:2018ybe}. In addition, $\mathcal{D}^2=0$ provided that the Heterotic Bianchi identity holds. 

The differential operators that appear in $\mathcal{D}$ are related to $\delb$, except for the terms which generate $\imath_{\mu}\del b$ in the action \eqref{eq:quadratic_action_full}. In fact, assuming the relevant $(2,0)$ cohomology vanishes, this can be absorbed in a redefinition of the other fields. To see how this works, let us Hodge decompose $b$ as
\begin{equation}
    b=\delb \gamma_{(0,1)}+\delb^\dagger\gamma_{(0,3)}\,,
\end{equation}
where, as always, we are ignoring any harmonic modes. The term $\delb \gamma_{(0,1)}$ can then be absorbed by a redefinition of $x$, while the remaining $\delb^\dagger\gamma_{(0,3)}$ term couples only to the $\delb$-exact part of $\chi=\imath_{\mu}\Omega$.\footnote{This observation requires us to note that $\del$ anti-commutes with $\delb^\dagger$ when acting on purely anti-holomorphic forms. This is also true in the non-K\"ahler case.} By an elementary integration by parts, this term can be absorbed in a redefinition of $\kappa=k\,\Omega$.

In the new field basis, the full action can be written as
\begin{equation}
\label{eq:RedefAction}
    S \equiv S_2 + S_2^{\text{gauge}}= \int_{X} \langle  \hat{y},\bar{D}\hat{y} \rangle _{Q}\wedge \Omega + \delb b \wedge \kappa \, ,
\end{equation}
where $\hat{y}\in \Omega^{0,1}(Q)$, $\bar{D}\colon\Omega^{0,p}(Q)\rightarrow \Omega^{0,p+1}(Q)$, and $\langle \cdot,\cdot\rangle _{Q}$ is a pairing on $\Omega^{0,\bullet}(Q)$. We see that with this redefinition, the action decomposes into a term involving the differential structure $\bar{D}$ found in \cite{delaOssa:2014cia}, and a term coupling $b$ and $\kappa$. We give more details on the connection $\bar{D}$ in Appendix \ref{app:FluxOperator}. Crucially, even with arbitrary flux, $\bar{D}$ can be interpreted as a Chern connection on the bundle $Q$ viewed as an extension bundle defined by the fluxes $(H,F)$~\cite{delaOssa:2017gjq, Garcia-Fernandez:2023vah, Garcia-Fernandez:2023nil}. This means that the absolute value of the partition function takes the form
\begin{equation}
\label{eq:FluxPartitionFunc}
    \vert Z_{\text{free}}\vert =\frac{I(Q)^{1/2}}{I(\Lambda^{0,0})}\,,
\end{equation}
which is precisely what we found in \eqref{eq:Z_free_RS_torsions}, except that $Q$ now includes an $\End(V)$ summand. The reader may wonder if the non-local field redefinitions we have used could introduce extra Jacobian factors in the path integral. This will not be the case, as we have simply shifted the fields by appropriate translations. 
We provide further discussion of such field redefinitions in a wider class of holomorphic field theories in section~\ref{sec:hol-field-redef}. 
Finally, we note that the form of \eqref{eq:FluxPartitionFunc} also holds when the various cohomologies that we have assumed to be trivial do not vanish, as long as the definition of the analytic torsion is extended to include the volumes of the relevant cohomologies.

\section{Anomalies}\label{sec:anomaly_review}

At this point, we have introduced a heterotic version of Kodaira--Spencer theory, derived from the heterotic superpotential, which controls deformations of the Hull--Strominger system (or, more precisely, F-term deformations). We have discussed how to define the one-loop partition function of this quasi-topological field theory and computed it for both $H=0$, and non-vanishing gauge fields.
In the next three sections, we discuss the anomalies that may be present in this theory and their resolutions. To make full use of the mathematical machinery surrounding anomalies, we will from here onward assume the existence of a background Kähler metric, even though the geometry may be torsional and the flux need not vanish. This background metric can be thought of as the first term in an $\alpha'$ expansion of the metric. We will focus on gauge and gravitational anomalies in Section \ref{sec:gauge_section}, and an anomaly under changes of a background metric in Section \ref{sec:metric_anomaly}. That is, a non-topological dependence of the partition function on a choice of background metric. First, however, we give a general discussion of anomalies for field theories on non-trivial backgrounds.

For the purposes of this paper, it will be useful to distinguish between two different types of anomalies. The first of these are conventional gauge anomalies which are captured by anomaly polynomials. Given a classical gauge-invariant action, the presence of these anomalies indicates that the partition function of a theory is not gauge invariant, and so the quantum theory is inconsistent. An example of this is a theory with charged chiral fermions where both axial and vector $\Uni 1$'s are gauged. The second kind of anomaly that we will come across is when a quantum theory depends on more data than the classical theory would suggest. An example of this is Chern--Simons theory, where the theory is classically topological -- independent of a choice of metric -- but the partition function of the theory is not; instead, it depends on a choice of framing for the three-manifold on which the theory is placed~\cite{Witten:1988hf}. Despite this, the partition function has a well-defined behaviour under changes of framing; thus, the quantum theory is consistent and should be thought of as computing invariants of framed three-manifolds. These anomalies are closer in spirit to `t Hooft anomalies for a global symmetry. Upon coupling the current for a global symmetry with a `t Hooft anomaly to a background gauge field, the partition function -- which now depends on a choice of background field -- will not be invariant under gauge transformations of the background field. This is an obstruction to gauging the global symmetry; one wants gauge-equivalent field configurations to contribute equally to the path integral, however the `t Hooft anomaly implies that this is not the case. In the case of Chern--Simons theory, the classical theory does not depend on a choice of metric for the three-manifold, but the quantum theory does depend on a background metric via a choice of framing. If one tried to `gauge' the classical topological symmetry by summing over only topologically inequivalent geometries, this would be inconsistent due to the transformation of the partition function under changes of framing. It is in this sense that Chern--Simons theory has a `t Hooft anomaly for its classical topological symmetry.

The holomorphic anomaly of Kodaira--Spencer theory, discussed at length in~\cite{Bershadsky:1993ta,Bershadsky:1993cx}, is similar. This anomaly comes from that fact that the action of the theory depends only on the holomorphic data of the complex structure moduli space, but the partition function actually has an anomalous dependence on the anti-holomorphic data, and so it is not a holomorphic function on the space of complex structures. In addition, since the theory is defined on a six-manifold, one might also find conventional gauge anomalies, such as the non-invariance of the partition function under gauge transformations of fluctuations of the complex structure. (Though Kodaira--Spencer has no chiral fermions, holomorphic field theories behave in many respects like chiral theories.) We shall return to this question later and see that while Kodaira--Spencer theory has a gauge anomaly, the honest B-model does not.

The holomorphic field theory that we are studying is in many respects a generalisation of Kodaira--Spencer coupled to holomorphic Chern--Simons, as it describes the fluctuations of a complex structure coupled to a hermitian metric and a gauge field. As is the case for Kodaira--Spencer theory, we might expect to find both kinds of anomalies outlined above. In other words, the corresponding quantum theory may depend on more parameters of the on-shell background about which we expand than the classical theory, signalling the presence of a topological or holomorphic anomaly. Furthermore, the theory itself may not be well-defined due to a gauge or gravitational anomaly.\footnote{As a note of caution, like KS theory, the theory we consider is power-counting non-renormalizable. This means that gauge anomalies (and counter-terms) are not restricted to one-loop and can appear at any order in the loop expansion. In previous examples, cancelling the one-loop anomalies has been enough to ensure cancellation to all loop orders~\cite{1201.4501,Costello:2015xsa,Bittleston:2022nfr}, so we are hopeful that the same will be true here.} The calculation of these anomalies is somewhat distinct, so we shall go through each in turn.

First, an anomaly indicating that the theory depends on more parameters than initially thought is in principle checked by computing the partition function of the theory. In practice, this is a difficult (often impossible) task to accomplish to all-loop order. Instead, we focus on computing the one-loop partition function, which depends on the quadratic part of the action. The formal answer for the partition function is then given in terms of determinants of operators which appear in the kinetic terms of the theory, with the determinants regularised in a gauge-invariant manner. As we saw in Section \ref{sec:one-loop}, one usually computes only the absolute value of the determinants, with the phase determined by a generalisation of the $\eta$-invariant. One can then study whether the absolute value of the one-loop partition function depends on more parameters than the classical action would suggest. In this paper, we focus on whether the modulus of the partition function remains independent of the background hermitian metric used for quantisation. We often refer to this as a ``metric anomaly''.

Second, and more importantly, we should also ask whether the theory has any gauge or gravitational anomalies. If these are present, the quantum theory is inconsistent. Generally, a partition function $Z$ will depend on some parameters and should be interpreted as a section of a certain determinant line bundle over the configuration space of on-shell backgrounds.
Let us denote this bundle by $\text{Det}$. In practice, the absolute value of the partition function $|Z|$ is given by ratios of determinants of elliptic operators. This is an infinite sum over the eigenvalues of the relevant operators and, using a gauge-invariant regulator, will automatically be gauge invariant. The phase of the partition function has no such guarantee. Though one often cannot compute the phase of $Z$ explicitly, one \emph{can} check whether or not the phase is gauge invariant.

Since $|Z|$ is gauge invariant, the bundle $\text{Det}$ has a natural hermitian metric and so an associated Chern connection. If this connection has non-zero curvature $\mathcal{F}_{\text{Det}}$, the phase of $Z$ will not come back to itself when one traverses a closed loop in the space of on-shell backgrounds. The phase of $Z$ then does not depend only on the gauge-invariant data of a point in the configuration space (a gauge orbit) and is in fact gauge variant, indicating a gauge anomaly. In our case, one might wonder whether such a change in phase is fatal given the discussion around \eqref{eq:phase_change}. However, the point is that once we pick a convention for the phase, it should be well defined under a change of gauge. In the case of a smooth base manifold $X$ where the kinetic terms in the action are Dirac operators, a formula for the curvature $\mathcal{F}_{\text{Det}}$ was given by Bismut--Freed in \cite[Theorem 1.21]{BismutFreed1,Bismut:1986wr}. In the case where $X$ is K\"ahler and the kinetic operators are complex differentials, the corresponding formula was given by Bismut--Gillet--Soul\'e in \cite[Theorem 1.27]{bismut1988analytic1,bismut1988analytic2,bismut1988analytic3}. Let us see how this works for the examples of holomorphic Chern--Simons theory~\cite{Witten:1992fb,1201.4501,Costello:2015xsa}, Kodaira--Spencer and the B-model on a Calabi--Yau threefold $X$~\cite{Bershadsky:1993cx}.\footnote{For a nice review of this material, see \cite{Costello:2019jsy}.}

The action of holomorphic Chern--Simons theory on a complex three-fold $X$ with $c_1(X)=0$ and a hermitian vector bundle $E$ with gauge group $G$ is
\begin{equation}
S[A]=\int_{X}\Omega\wedge\tr\left(A\wedge\bar{\partial}A+\tfrac{2}{3}A\wedge A\wedge A\right)\, ,\label{eq:HCS_action}
\end{equation}
where $A\in\Omega^{0,1}(\End(E))$ is a Lie algebra-valued $(0,1)$-form on $X$ with respect to the complex structure associated to the holomorphic three-form $\Omega$. The local invariances of this theory include gauge transformations $\bar{\partial}\to k^{-1}\bar{\partial}k$, where $k$ is a complex gauge transformation of the frame for the vector bundle $E$, and local (complex) coordinate changes that preserve $\Omega$. In particular, the action is invariant under the infinitesimal gauge transformation
\begin{equation}
\delta A=\bar{\partial}c+[A,c]\, ,\qquad c\in\Omega^{0}(\End(E))\, .
\end{equation}
The equations of motion that follow from the action are simply
\begin{equation}\label{eq:holCS_eom}
\bar{\partial}A+A\wedge A=0\, .
\end{equation}
This says that $\bar{\partial}+A$ defines a new $(0,1)$ connection whose curvature has no $(0,2)$ component, which is equivalent to a holomorphic structure on vector bundles associated to $E$. Solutions to the linearised equations of motion modulo the gauge freedom are inequivalent holomorphic structures on $E$, parametrised by elements of $H^{1}(X,\End(E))$. 

This theory is most easily quantised in the Batalin--Vilkovisky (BV) formalism~\cite{Batalin:1981jr} by adding a ghost field in $\Omega^{0,0}(\End(E))$, and antifields for both $A$ and the ghost field. These can be collected in a single field $\mathcal{A}$ living in $\Omega^{0,\bullet}(\End(E))$, where the component in $\Omega^{0,p}(\End(E))$ has ghost number $1-p$. As is the case for conventional Chern--Simons theory~\cite{hep-th/9110056}, the BV action takes the same form as \eqref{eq:HCS_action} with $A$ replaced by $\mathcal{A}$, and gauge fixing the action requires the introduction of a hermitian metric on $X$ compatible with the complex structure defined by $\Omega$.

Upon quantising this theory around an on-shell configuration $A_{0}$ which solves \eqref{eq:holCS_eom}, one finds that the absolute value of the one-loop partition function simplifies to~\cite{thomas1997gauge}
\begin{equation}
|Z|^2=I(\Lambda^{0,0}(\End(E)))\, ,
\end{equation}
where the relevant Laplacian is defined by the differential $\bar{D}=\bar{\partial}+[A_{0},\cdot]$ and $I(\Lambda^{p,0}(\End(E)))$ is the Ray--Singer analytic torsion for $\bar{D}$ acting on a complex of $\End(E)$-valued $(p,0)$-forms~\cite{Ray:1973sb}. This answer will depend on the classical data of $\Omega$ (which also defines a complex structure) and the hermitian metric used in gauge fixing the BV action. In particular, though the classical action \eqref{eq:HCS_action} is quasi-topological (depending only on the complex structure and gauge field), the quantum theory also depends on the choice of background metric on $X$. This is the analogue of the topological anomaly seen in conventional Chern--Simons. In principle, the phase of $Z$ may (and is likely to) depend on the metric as well via a generalisation of the $\eta$-invariant, though this is more difficult to understand.\footnote{For a discussion of how this works for four-folds, see \cite{thomas1997gauge}.}

Now we ask whether the theory has any gauge anomalies. The partition function $Z$ should be interpreted as a section of a certain holomorphic determinant line bundle $\text{Det}$ over the (complex) configuration space $\mathcal{M}$ of holomorphic structures on $E$ and holomorphic three-forms on $X$. In order to appeal to the results of \cite{bismut1988analytic1,bismut1988analytic2,bismut1988analytic3}, we assume that the background metric on $X$ is actually K\"ahler.
If the Chern connection associated to $|Z|$ has non-zero curvature $\mathcal{F}_{\text{Det}}$, the phase of $Z$ does not depend only on the gauge-invariant data of a point in the configuration space (a gauge orbit in the space of holomorphic structures or holomorphic three-forms).
The phase of $Z$ is thus not gauge invariant, indicating a gauge anomaly. In our case, the expression for the curvature of $\text{Det}$ is given by\footnote{As discussed in \cite{Bittleston:2022nfr}, the anomaly polynomial is actually defined in terms of characteristic classes of bundles over the total space of $\mathcal{X}\to\mathcal{M}$, where $\mathcal{M}$ is the configuration space of the on-shell background and the fibres of $\mathcal{X}$ are $X$.}
\begin{equation}
\FDet=\frac{1}{2}\left[\int_{X}\td(X)\wedge\ch(\End(E))\right]_{(1,1)}\,,
\end{equation}
where $\td$ and $\ch$ are the total Todd class and Chern character, and one picks out the $(4,4)$-component of the integrand to give a $(1,1)$-form on $\mathcal{M}$ after integration.
Expanding this out and using $c_{1}(X)=0$ (since $X$ admits a nowhere-vanishing three-form $\Omega$), one finds
\begin{equation}
\FDet=\frac{1}{2}\left[\int_{X}\op{ch}_{4}(\End(E))+\frac{1}{20}\dim G\op{ch}_{4}(X)-\frac{1}{12}\op{ch}_{2}(X)\wedge\op{ch}_{2}(\End(E))\right]_{(1,1)}.
\end{equation}
The first term is the pure gauge anomaly for holomorphic Chern--Simons theory. The second term is the gravitational anomaly associated to local coordinate changes that preserve $\Omega$. The final term is a mixed gauge-gravitational anomaly. Since in holomorphic Chern--Simons theory it is the gauge field $A$ that is dynamical, while $\Omega$ is viewed as a classical background field, the second and third terms should be interpreted as 't Hooft anomalies which are not fatal for the theory.\footnote{If one tried to integrate over choices of $\Omega$ up to gauge equivalence, the 't Hooft anomaly would be promoted to a gravitational anomaly for $\Omega$-preserving diffeomorphisms.} The problematic term is the first one. Writing this in terms of the background field $A_{0}$ as
\begin{equation}
\frac{1}{2}\int_{X}\op{ch}_{4}(\End(E))=\frac{1}{2}\frac{1}{4!}\left(\frac{\ii}{2\pi}\right)^{4}\int_{X}\tr(\partial A_{0})^{4}\,,
\end{equation}
the usual descent procedure implies that the one-loop box diagram with only the gauge field appearing is not gauge invariant. The gauge transformation of this diagram is then proportional to
\begin{equation}
\int_{X}\tr\left(c(\partial A)^{3}\right)\,,
\end{equation}
in agreement with the gauge anomaly found in \cite{1201.4501,Costello:2015xsa,Costello:2019jsy}.

Similarly, we can use the formula for $\mathcal{F}_{\text{Det}}$ to compute the anomalies of the theories that come from the $SL(3,\bC)$ and generalised Hitchin functionals on $X$~\cite{Pestun-Witten2005}. At one-loop, the latter agrees with the topological B-model on a K\"ahler manifold with $c_1(X)=0$~\cite{Bershadsky:1993cx}. In both cases, the one-loop partition function takes the form
\begin{equation}
    |Z| = \frac{I(\Lambda^{1,0})}{I(\Lambda^{0,0})^{\alpha}}\, ,
\end{equation}
where $\alpha=1$ for the $SL(3,\bC)$ theory and $\alpha=3$ for the generalised Hitchin / B-model~\cite{Bershadsky:1993cx,Pestun-Witten2005}. The curvature on Det is then
\begin{equation}
    \FDet = \frac{1}{2}\left[\int_{X} \td(X)\wedge \ch(\Omega^{1,0}) - \alpha\td(X)\right]_{(1,1)}\, .
\end{equation}
Expanding this out using $c_1(X)=0$ and the identities in \cite{sati}, one finds 
\begin{equation}\label{eq:KS_anomaly}
    \FDet = \frac{3-\alpha}{20}\int_X \ch_4(X)\, .
\end{equation}
This indicates the presence of a gravitational anomaly for $\Omega$-preserving diffeomorphisms. Unlike holomorphic Chern--Simons where $\Omega$ is viewed as a classical background field, the $SL(3,\bC)$ and B-model theories contain degrees of freedom which describe fluctuations of $\Omega$. Thus, \eqref{eq:KS_anomaly} is not a `t Hooft anomaly and is fatal if not cancelled. Fortunately, the curvature, and hence the gravitational anomaly, vanishes if $\alpha=3$ but not generically for $\alpha=1$. This is consistent with the fact that the B-model is a well-defined quantum theory on Calabi--Yau manifolds, while the $SL(3,\bC)$ theory alone is not.\footnote{If one does not assume $c_1=0$, one finds that even for $\alpha=3$, $\mathcal{F}_{\text{Det}}$ receives contributions from terms of the form $c_1^2 c_2$ and $c_1 c_3$, indicating that the B-model is not consistent when $c_1\neq 0$.}

In the next section, we analyse the gauge, gravitational and mixed anomalies for our holomorphic field theory. Following this, in Section \ref{sec:metric_anomaly}, we study how the one-loop partition function depends on the choice of the background metric introduced by quantisation. We will leave the study of a potential holomorphic anomaly to future work.

\section{Gauge and gravitational anomalies}\label{sec:gauge_section}

Our theory is classically invariant under gauge transformations by $\mathcal{D}$-exact shifts of the form \eqref{eq:gauge_D_def}. These gauge invariances may not persist in the quantum theory, indicating the presence of an anomaly. Unlike a holomorphic or topological anomaly, these gauge and gravitational anomalies render the theory ill-defined if they are not cancelled.

Given the form of the partition function in \eqref{eq:FluxPartitionFunc}, the curvature of the connection on $\text{Det}$ which captures anomalies is given by
\begin{equation}\label{eq:Det_curv}
\FDet=\frac{1}{2}\left[\int_{X}\td(X)\wedge\ch(Q)\right]_{(1,1)}-\left[\int_{X}\td(X)\right]_{(1,1)}\, ,
\end{equation}
where we have used $\ch(\Lambda^{0,0})=1$ in the second term.\footnote{Recall that, formally, $Q$ and $TX$ are extended to be bundles over the total space formed by fibring $X$ over the moduli space. Thus, for example, $\ch(Q)$ can have an eight-form component, but $c_r(Q)=0$ for $r>\rk Q$.} It should be noted that the last term in \eqref{eq:Det_curv}, coming from the coupling between the off-shell modes $b$ and $\kappa$ in \eqref{eq:RedefAction}, will be important when we come to factorise the anomaly polynomial and the subsequent cancellation of certain anomalies.

Focusing on the case where $X$ is Calabi--Yau, recall that the Todd class of $X$ simplifies for $c_1(X)=0$ to
\begin{equation}
\begin{aligned}
    \td(X)&=1+\tfrac{1}{12}c_2(X) + \tfrac{1}{240}c_2(X)^2+\dots = 1-\tfrac{1}{12}\ch_2(X)+\tfrac{1}{20}\ch_4(X)+\dots
\end{aligned}
\end{equation}
where $c_4(X)$ and higher vanish as $TX$ is a rank-three bundle, and we do not need any further terms, since these are sufficient to fix the $(1,1)$-component of \eqref{eq:Det_curv} on the moduli space. Similarly, the Chern character of $Q$ can be expanded as
\begin{equation}
    \ch(Q) = \rk (Q) +\ch_1(Q)+\ch_2(Q)+\ch_3(Q)+\ch_4(Q)+\dots
\end{equation}
Putting these together, $\mathcal{F}_{\text{Det}}$ simplifies to
\begin{equation}
\FDet=\frac{1}{2}\left[\int_{X}\ch_{4}(Q)+\frac{1}{20}\bigl(\rk(Q)-2\bigr)\int_{X}\ch_{4}(X)-\frac{1}{12}\int_{X}\ch_{2}(X)\wedge\ch_{2}(Q)\right]_{(1,1)}\, .
\end{equation}
Finally, since we assume $X$ is Calabi--Yau, the curvature of the Chern connection on $TX$ transforms in the fundamental representation of $\SU3$, which implies that the four- and eight-form components of the Chern character are related by
\begin{equation}
    \ch_4(X) = \frac{1}{12} \ch_2(X)^2\, .
\end{equation}
Thus, on a Calabi--Yau, the anomaly of our heterotic version of Kodaira--Spencer theory is determined by
\begin{equation}\label{eq:final_CY_anomaly}
\FDet=\frac{1}{2}\left[\int_{X}\ch_{4}(Q)+\frac{1}{240}(\rk(Q)-2)\int_{X} \ch_2(X)^2-\frac{1}{12}\int_{X}\ch_{2}(X)\wedge\ch_{2}(Q)\right]_{(1,1)}\, .
\end{equation}
In the remainder of this section, we analyse how this anomaly constrains the background geometry and gauge sector.

\subsection{Special case: \texorpdfstring{$Q=T^{1,0}\oplus T^{*1,0}$}{Q = T10 + T*01}}
\label{sec:CYanomaly}
We first consider the special case where, in addition to $X$ being Calabi--Yau, the gauge bundle is trivial, so that $Q$ splits as a direct sum of bundles, $Q=T^{1,0}\oplus T^{*1,0}$. In this case, the Chern character of $Q$ is related to that of $X$ by
\begin{equation}
    \ch_n(Q) = 2 \ch_n(X)\, ,
\end{equation}
for $n$ even, while it vanishes for $n$ odd. The contribution to the curvature polynomial due to the Chern character reduces to just the rank, and the expression \eqref{eq:final_CY_anomaly} becomes
\begin{equation}
\FDet=\frac{1}{120}\left[\int_{X} \ch_2(X)^2\right]_{(1,1)}\, .
\end{equation}
In this simple case, we see there will naively be a gravitational anomaly unless $\ch_2(X)=0$. In fact, with a trivial gauge bundle and no $H$-flux, this is equivalent to the Bianchi identity for $H$ restricted to $X$.

Through the descent procedure, a change in frame given by a local rotation $\gamma = \nabla v_{0}$, where $v_{0} \in \Gamma(T^{1,0}) \subset A^{0}$ is the vector component of the corresponding gauge parameter and $\nabla$ is the covariant derivative associated to the background metric $g$, gives a change in the phase of the partition function proportional to
\begin{equation}
    \int_{X} \ch_2(X) \tr(\cal R \gamma)\, .
\end{equation}
We have written ${\cal R} = \tfrac{\ii}{2\pi}R$ , where $R$ is the Ricci scalar associated to the background metric on $X$. This clearly vanishes if $\ch_{2}(X)=0$, but it can also be cancelled via an appropriate Green--Schwarz mechanism even if we do not assume the Bianchi identity. To do so, we include a counter-term in the Lagrangian of the form
\begin{equation}\label{eq:no_gauge_counter}
    S_{\text{counter}} = \frac{1}{240}\int_{X}\ch_2(X)\wedge x\, ,
\end{equation}
and give an anomalous gauge transformation to $x$:\footnote{Note that we are working in the redefined theory in which $\del b$ drops out of the action, and hence the gauge transformation for $x$ does not contain a $\del b_{0}$ term.}
\begin{equation}\label{eq:x_anomalous_transform}
\begin{aligned}
    \delta x &= \delb x_{0} - \hbar \tr({\cal R}\gamma) \\
    &= \delb x_{0} - \hbar\tr({\cal R} \, \nabla v_{0})\, ,
\end{aligned}
\end{equation}
where $x_{0} \in \Omega^{1,0}\subset A^{0}$ is the original gauge parameter for $x$. We have also included an explicit $\hbar$ in the gauge transformations to indicate that the anomalous gauge transformation is to a cancel one-loop effect.\footnote{To be precise, we expand a given field $\Phi$ in $\hbar$, where the leading term is classical, i.e.\ solves the equations of motion. Modulo higher-loop ${\cal O}(\hbar^2)$ effects, the classical action then remains invariant under the Green--Schwarz transformation.}

\subsection{Including the gauge fields}

We now reinstate the gauge sector. To study the possible anomalies associated to this, we note that gauge transformations of $A$ are equivalent to rotations of the frame for the vector bundle $V$. Thanks to this, we can collectively consider the gauge and gravitational anomalies as anomalies associated to frame rotations of $Q= T^{1,0}X\oplus \End(V)\oplus T^{*1,0}X$. We therefore have
\begin{equation}\label{eq:chern_decomp}
    \ch_{n}(Q) = 2\ch_{n}(X) + \ch_{n}(\End(V))
\end{equation}
for $n$ even. The expression \eqref{eq:final_CY_anomaly} for the curvature on $\text{Det}$ which determines the anomalies associated with such background rotations is now
\begin{equation}\label{eq:AnomalyPolGauge2}
    \FDet = \frac{1}{2}\biggl[\int_{X} \ch_{4}(\End(V)) - \frac{1}{12} \ch_{2}(X)\wedge \ch_{2}(\End(V)) + \frac{4+\dim G}{240} \ch_{2}(X)^{2} \biggr]_{(1,1)}\, .
\end{equation}
Note that one should be careful with this expression. If the gauge field flux $F$ is non-vanishing, then the bundle $Q$ is really an extension bundle with the various summands non-trivially twisted~\cite{Anderson:2014xha, delaOssa:2014cia,Garcia-Fernandez:2015hja}. %
The relation \eqref{eq:chern_decomp} then holds only at the level of cohomology, and hence the expression above holds only up to some $\dd$-exact terms. However, we need only check that we can twist the line bundle, of which $Z_{\text{free}}$ is a section, through the addition of counter terms to the action such that a flat connection exists. If this is the case, then there is vanishing holonomy and so $Z_\text{free}$ is constant around contractible loops.\footnote{There can be discrete holonomy of a flat bundle around non-contractible loops. The $\dd$-exact terms may then play a role in the study of such global anomalies, which we do not consider.} The only relevant information for cancelling local anomalies is therefore the cohomology class of $\FDet$.

With this in mind, the curvature \eqref{eq:AnomalyPolGauge2} does not vanish for generic choices of gauge group $G$. As in the holomorphic Chern--Simons case, the first term can be identified as a gauge anomaly, while the final term can be interpreted as a gravitational anomaly. Naively, we also expect a mixed anomaly from the middle term. This, however, vanishes in our case since the operator $\bar{D}$ on $Q$, which is used to define the Ray--Singer torsion $I(Q)$, satisfies\footnote{This and other properties of $\bar{D}$ are reviewed in Appendix \ref{app:FluxOperator}.}
\begin{equation}
    \bar{D}^{2} = 0 \qquad \Rightarrow \qquad \ch_{2}(\End(V)) = 0\, .
\end{equation}
Even with this, we still have gauge and gravitational anomalies to contend with, and so we are led to introduce a Green--Schwarz type mechanism to cancel them. To do so, we require the anomaly polynomial to factorise for which we need to constrain the form of the gauge group. The simplest constraints to put on the gauge group are
\begin{equation}\label{eq:chern_character_constraint}
    \ch_{4}(\End(V)) \propto \ch_{2}(\End(V))^{2}\, .
\end{equation}
A list of simple groups which satisfy the above constraint, along with useful trace identities, are listed in Appendix \ref{app:trace_identities}. The previous discussion implies that the gauge anomaly vanishes for these groups.

Given such a choice of gauge group, we are left with the gravitational anomaly to cancel. One might expect that we could use the same anomaly cancelling term as in the previous subsection. However, once the gauge sector is turned on, the flux $F$ appears in the gauge transformation of $x$ by $\mathcal{D}$-exact terms, as in \eqref{eq:gauge_D_def}, and so the counter-term given in \eqref{eq:no_gauge_counter} is no longer gauge invariant. Instead, we can use the fact that, locally, $\ch_{2}(X) \sim \del \omega_{\text{CS}}^{(1,2)}$ for some $(1,2)$-form which we can take to be gauge invariant and $\delb$-closed. We can then add the local counter-term\footnote{The anomaly in the previous section can also be cancelled by such a counter-term. However, only in the previous case can we cancel the anomaly with a global counter-term, as given in \eqref{eq:no_gauge_counter}.}
\begin{equation}\label{eq:simple_gauge_counter}
    S_{\text{counter}} = \frac{4+\dim G}{480}\int_{X} \omega_{\text{CS}}^{(1,2)} \wedge \chi\, ,
\end{equation}
provided we give $\chi$ an anomalous gauge transformation\footnote{The transformation of the two-form $x$ in \eqref{eq:x_anomalous_transform} is similar to the usual Green--Schwarz mechanism \cite{Green:1984sg}, while the transformation of $\chi$ resembles more the Green--Schwarz type transformations introduced when cancelling anomalies in holomorphic gauge theories coupled to gravity, see for example \cite{Costello:2015xsa, Williams:2018ows, Costello:2019jsy, Costello:2021bah, Costello:2021kiv, Costello:2022wso, Bittleston:2022nfr}.}
\begin{equation}
    \delta \chi = \delb \chi_{0} - \hbar\, \del \tr({\cal R}\gamma)\, ,
\end{equation}
where $\chi_{0} \in \Omega^{2,0} \subset A^{0}$ is the original gauge parameter of $\chi = \imath_{\mu}\Omega$. Of course, the counter-term \eqref{eq:simple_gauge_counter} is only locally well-defined in general. A more thorough treatment would require extending the theory to some $\tilde{X}$ which is bounded by $X$, as in the Green--Schwarz mechanism in, for example, six-dimensional supergravity~\cite{Monnier:2018nfs}. Such considerations can have interesting consequences for global anomalies. For the local anomaly, however, \eqref{eq:simple_gauge_counter} is sufficient.

\subsection{Aside -- a different, untwisted model}\label{sec:untwisted_model}

We pause for a moment to consider a slightly different model which arises from considering $Q= T^{1,0}\oplus \End(V)\oplus  T^{*1,0}$ as simply a direct sum of bundles, and not as an extension structure. This bundle comes equipped with a differential
\begin{equation}
\bar{D} =\begin{pmatrix}
\delb & 0 & 0\\
0 & \delb_A & 0\\
0 & 0 & \delb
\end{pmatrix}\, ,
\end{equation}
which squares to zero if and only if $X$ is a complex manifold and $\End(V)$ is a holomorphic bundle on $X$. In particular, nilpotency of $\bar D$ imposes no constraints on the Chern characters of $\End(V)$. We can then consider the theory as in \eqref{eq:RedefAction}, but with $\bar{D}$ as above. It is worth noting that this theory \emph{does not} arise from perturbations of the superpotential with generic gauge fields and so does not correspond to a Kodaira--Spencer-like theory. Instead, we view this as a toy model of a holomorphic field theory and then study its associated anomalies. Given that the Chern characters of the bundles are no longer constrained, we will instead use the heterotic Bianchi identity to restrict anomaly cancellation.

The curvature which determines the anomalies is again given by \eqref{eq:AnomalyPolGauge2}, but now the mixed anomaly does not cancel automatically. Instead, it must also be cancelled by a Green--Schwarz mechanism. To do so, we require that the gauge group is such that \eqref{eq:AnomalyPolGauge2} factorises. This is most easily done by again imposing
\begin{equation}
    \ch_{4}(\End(V)) \propto \ch_{2}(\End(V))^{2}\, .
\end{equation}
A quick calculation shows that if $G$ is simple then, even with the above constraint, the anomaly polynomial does not factorise. However, we can make progress if we also impose the heterotic Bianchi identity. We demonstrate this with an example.

\subsubsection*{$G=SO(8)$}

For the gauge group $G=SO(8)$, the identities 
\begin{equation}
    \ch_{4}(\End(V)) = \tfrac{1}{2}\ch_{2}(V)^{2} \, , \qquad \ch_{2}(\End(V)) = 6\ch_{2}(V)\, ,
\end{equation}
which are derived from the formulae in Appendix \ref{app:trace_identities}, can be used to rewrite the curvature of Det as
\begin{equation}
    \FDet = \frac{1}{2}\left[\int_X \frac{1}{2}\ch_{2}(V)^{2} - \frac{1}{2}\ch_{2}(X)\wedge \ch_{2}(V) + \frac{2}{15}\ch_{2}(X)^{2} \right]_{(1,1)}\, .
\end{equation}
It is clear that this does not factorise as it stands. However, the heterotic Bianchi identity is equivalent to the identity
\begin{equation}
    \ch_{2}(X) = \ch_{2}(V)\, ,
\end{equation}
which holds in cohomology. Ignoring any exact terms, as in the discussion after Equation \eqref{eq:AnomalyPolGauge2}, we have
\begin{equation}
    \FDet = \frac{1}{15}\left[ \int_{X} \ch_{2}(X)^{2} \right]_{(1,1)}\, .
\end{equation}
We can then cancel this anomaly using the counter-term in \eqref{eq:no_gauge_counter}, exactly as in the case without gauge fields. This counter-term is gauge invariant in this case, since the form of $\bar{D}$ implies that the fluxes do not appear in the gauge transformation of $x$.

This procedure can be used to cancel the gauge, gravitational and mixed anomalies for any simple group $G$ satisfying \eqref{eq:chern_character_constraint}. A natural extension is to consider product groups of the form $G=G_{1}\times G_{2}$, with each $G_{i}$ simple. Remarkably, we find that any combination of groups satisfying \eqref{eq:chern_character_constraint} leads to factorisation of the anomaly polynomial. A full list of such factorisations is given in Appendix \ref{app:trace_identities}. We shall give one example of this, as the procedure holds, mutatis mutandis, for the other cases.

\subsubsection*{$G=SU(3)\times E_6$}
As an example, we consider the case where $V$ is an $SU(3)\times E_6$ bundle. Using the trace identities found in Appendix \ref{app:trace_identities}, the curvature on Det becomes\footnote{To go from the first to the second line, we have moved from $\ch_{2}(\End(V)) = \tfrac{1}{2}\Tr({\cal F}^{2})$ to $\ch_{2}(V) = \tfrac{1}{2} \tr({\cal F}^{2})$, where $\Tr$ and $\tr$ denote the trace in the adjoint and fundamental representations respectively. The relation between these two traces is given in Appendix \ref{app:trace_identities} for various groups. } 
\begin{equation}\label{eq:GaugeAnomalySU3E8}
    \begin{aligned}
        \FDet &= \frac{1}{2} \left[ \int_{X} \ch_{4}(\End(V_{1})) - \frac{1}{12}\ch_{2}(X)\wedge\ch_{2}(\End(V_{1})) + \frac{3}{8}\ch_{2}(X)^{2} \right. \\
        & \qquad \qquad \left. +\ch_{4}(\End(V_{2})) - \frac{1}{12}\ch_{2}(X)\wedge \ch_{2}(\End(V_{2})) \right]_{(1,1)} \\
        &= \frac{1}{2}\left[\int_{X} \frac{1}{24}\left(\ch_{2}(X)-6\ch_{2}(V_{1})\right)^{2} + \frac{1}{12}\left(2\ch_{2}(X) - \ch_{2}(V_{2})\right)^{2} \right]_{(1,1)}\, .
    \end{aligned}
\end{equation}
Via the descent procedure, the anomalous change in the phase of the partition function is proportional to
\begin{equation}
\begin{aligned}
    &\int_{X}\frac{1}{48}\bigl(\ch_{2}(X) - 6\ch_{2}(V_{1})\bigr)\bigl(\tr({\cal R}\gamma) - 6\tr({\cal F}_{1}\epsilon_{1})\bigr) \\
    &\qquad +\int_{X} \frac{1}{24}\bigl(2\ch_{2}(X)-\ch_{2}(V_{2})\bigr) \bigl(2\tr({\cal R}\gamma) - \tr({\cal F}_{2}\epsilon_{2})\bigr) \, ,
\end{aligned}
\end{equation}
where ${\cal R} = \tfrac{\ii}{2\pi}R$ and  ${\cal F}_{i} = \tfrac{\ii}{2\pi}F_{i}$, and we have set $\delta {\cal R}=\gamma$ and $\delta {\cal F}_i=\epsilon_i$, where $\gamma$ is an infinitesimal frame rotation for $T^{1,0}$ and $\epsilon_i$,  $i=1,2$, are infinitesimal background gauge transformations. 

To cancel these anomalies, we assume that one of the two terms is trivial in cohomology. For example, we can assume that, as cohomology classes, we have
\begin{equation}
        2 \ch_{2}(X) =\ch_2(V_2)\,.
\end{equation}
This trivialises the second term in $\FDet$, and so it can be neglected for local anomaly cancellation. The anomaly that arises from the other term can then be cancelled by the counter-term
\begin{equation}\label{eq:semi-simple_gauge_counter}
    S_{\text{counter}} = \frac{1}{48}\int_{X} \bigl(\ch_{2}(X) - 6\ch_{2}(V_{1})\bigr)\wedge x\, ,
\end{equation}
together with an anomalous gauge transformation for $x$:
\begin{equation}
    \delta x = \delb x_{0} - \frac{\hbar}{48} \bigl(\tr ({\cal R}\gamma) - 6 \tr({\cal F}_{1}\epsilon_{1}) \bigr)\, .
\end{equation}

We point out that cancelling the anomaly required a triviality constraint on the second Chern character of the bundles involved, though the choice of constraint was somewhat arbitrary. For example, we could instead impose
\begin{equation}
    \ch_2( X )=6\ch_2(V_1)\, ,
\end{equation}
and then construct a counter-term containing $2\ch_{2}(X)-\ch_{2}(V_{2})$, with an appropriate anomalous transformation for $x$. Other triviality constraints could also be considered, leading to different choices of counter-terms and anomalous transformations of the fields.

In particular, it is interesting to note that for some choices of gauge bundle, namely $SO(8)\times SO(8)$ and $SU(3)\times E_6$, we need to assume only the triviality of the ten-dimensional anomaly constraint / Bianchi identity for $H$ in order to cancel these anomalies. This imposes a linear relationship between $\ch_{2}(X)$ and $ \ch_{2}(V_{i})$, which are, at the level of cohomology%
\footnote{The difference in coefficients comes from the fact that we want $\tfrac{1}{2}\tr{\cal R}^{2} = \hat{\tr}{\cal F}_{\text{total}}^2$, where $\hat{\tr}$ is the inner product normalised such that the long roots have weight two. The relative factor between that and the trace $\tr$ in the fundamental representation is given by the coefficients in the constraints.}
\begin{equation}
    \begin{aligned}
        SU(3)\times E_{6}\colon&&\tfrac{1}{2}\ch_{2}(X) - \ch_{2}(V_{1}) - \tfrac{1}{6} \ch_{2}(V_{2}) &= 0\, , \\
        SO(8)\times SO(8)\colon & &\tfrac{1}{2}\ch_{2}(X) - \tfrac{1}{2}\ch_{2}(V_{1}) - \tfrac{1}{2}\ch_{2}(V_{2}) &= 0\, .
    \end{aligned}
\end{equation}
Using this for $G=SU(3)\times E_6$, the curvature simplifies, up to $\dd$-exact terms, to
\begin{equation}
    \FDet = \frac{1}{16}\left[ \int_{X} \bigl(2\ch_{2}(X)-\ch_{2}(V_{1})\bigr)^{2} \right]_{(1,1)}\, ,
\end{equation}
which can again be cancelled with a counter-term exactly as in \eqref{eq:semi-simple_gauge_counter}.
It is interesting to note that if we identify the $SU(3)$ part of the gauge group with the structure group of the Calabi--Yau, as is done in the standard embedding \cite{Candelas:1985en}, then this is also consistent as a six-dimensional quantum theory.

\section{A metric anomaly}
\label{sec:metric_anomaly}

As we mentioned in Section \ref{sec:het_KS}, the quadratic action derived from the superpotential is manifestly independent of the hermitian metric on $X$. However, when computing the partition function, we were forced to introduce a background metric in order to define the adjoint operators, the Laplace operators and ultimately the analytic torsion. This can introduce an anomalous dependence on the background metric in the quantum theory, similar to that discussed in \cite{Witten:1988hf} for Chern--Simons theory and \cite{Pestun:2005rp} for the generalised Hitchin functional. The holomorphic anomaly discussed in \cite{Bershadsky:1993ta,Bershadsky:1993cx} is also of this type -- the classical theory depends only on the holomorphic data of the complex structure, $\mu$, but the partition function picks up an anomalous dependence on anti-holomorphic data, $\bar\mu$. In our case, we will refer to this as a ``metric anomaly'' to distinguish it from the previous section's gravitational anomaly (which signalled loss of invariance under local coordinate changes). Unlike the local anomalies analysed in the previous section, the presence of this metric anomaly does not render the theory inconsistent, but instead may break the quasi-topological nature of the partition function. That is, though the classical quadratic action appears to depend on only the complex structure data on $X$, the quantised theory may depend also on the choice of background metric. Interestingly, given certain topological constraints, we will find that we can cancel this metric anomaly.

In principle, one can check whether a metric anomaly is present by computing the partition function and then examining whether it depends on a choice of background metric. In practice, one cannot compute the partition function in generality. Instead, we will appeal to the analysis of Bismut et al.~\cite{bismut1988analytic1, bismut1988analytic2, bismut1988analytic3} which describes how the analytic torsion varies under a change of metric on a K\"ahler manifold (and a change of hermitian metric on a vector bundle over it). This will allow us to check whether the absolute value of the 
 one-loop partition function -- expressed in terms of Ray--Singer torsions -- depends on these background fields, and to understand whether the anomalous variation can be cancelled by local counter-terms.

To begin, we recall that the variation of the analytic torsion of a holomorphic bundle $V$ under a change in the K\"ahler metric $g$ on the manifold and the hermitian metric $h$ on the fibres of $V$ takes a similar form to the curvature associated to the gauge and gravitational anomalies \cite[Theorem 1.22]{bismut1988analytic1, bismut1988analytic2, bismut1988analytic3}:\footnote{Once again, we ignore the zero-modes. A more detailed analysis of the variation including the contribution from zero-modes can be found in \cite{Mueller1999ExtremalKM}. Also, as per our comment at the beginning of Section \ref{sec:anomaly_review}, we will assume that we are working around a Calabi--Yau metric $g$, at least to leading order in $\alpha'$.}
\begin{equation}
\label{eq:VarTorsion}
    \frac{1}{2\pi} \frac{\partial}{\partial t} \log I(V) = \frac12 \int_X \partial_t \left[ \td \left(\frac{1}{2\pi} (\ii R+t\,g^{-1}\delta g)\right)\wedge \ch\left(\frac{1}{2\pi} (\ii F+t\, h^{-1}\delta h)\right) \right]_{4}    \, ,
\end{equation}
where $R$ and $F$ are the curvatures of the Chern connections on the holomorphic tangent bundle and $V$ respectively, and the total Todd class and Chern character are expanded in terms of the indicated arguments. The subscript $4$ means that we take the component that is degree 4 in the curvatures $R,F$ and their variations.

For a completely arbitrary complex three-dimensional background $X$, we can express the Todd classes in terms of the Chern classes via%
\begin{equation}
    \begin{aligned} 
        \td(X)_0&=1,\\
\td(X)_1&=\frac12 \ch_1(X),\\
\td(X)_2&=\frac{1}{8}\ch_1(X)^2-\frac{1}{12}\ch_2(X),\\
\td(X)_3&=\frac{1}{48} \left( \ch_1(X)^3-2\ch_2(X)\ch_1(X)\right), \\
\td(X)_4&= \frac{1}{1440} \Bigl(  5\ch_1(X)^4 -30 \ch_2(X)\ch_1(X)^2 +60 \ch_3(X)\ch_1(X)\\
&\eqspace\phantom{\frac{1}{1440}\Bigl(}+20 \ch_2(X)^2 - 168 \ch_4(X)  \Bigr),
    \end{aligned}
\end{equation}
where $\ch_k(X)=\frac{1}{k!}\tr({\cal R}^k)$ and $\mathcal{R}=\frac{\ii}{2\pi}R$. Therefore, we find
\begin{equation}
\begin{aligned}
    \left[\td(X)\ch(Q)\right]_4&= \frac{\rk(Q)}{1440} \Bigl( 5\ch_1(X)^4 -30 \ch_2(X)\ch_1(X)^2 +60 \ch_3(X)\ch_1(X)\\
&\eqspace\phantom{\frac{\rk(Q)}{1440}\Bigl(} +20 \ch_2(X)^2 - 168 \ch_4(X)\Bigr)+\frac{1}{48} \ch_1(X)^3 \ch_1(Q)\\
&\eqspace - \frac{1}{24}\ch_2(X)\ch_1(X) \ch_1(Q)  +\frac{1}{8}\ch_1(X)^2\ch_2(Q)    \\
    & \eqspace-\frac{1}{12}\ch_2(X)\ch_2(Q)+\frac{1}{2} \ch_1(X)\ch_3(Q)+ \ch_4(Q)\,.
\end{aligned}
\end{equation}
There are further simplifications to this polynomial when we remember that $Q$ is defined as an extension. In particular, since the groups we consider are semisimple and compact, the bundles $\End(V)$ and hence the bundle $Q$ are self-dual, we can drop all traces of odd-powers of the curvature $\mathcal{F}$. Note also that $\mathcal{R}$ and $\mathcal{F}$ are trace-less, though their variations may not be. Hence, we can drop any powers of $\ch_1(X)$ and $\ch_1(Q)$ beyond linear order. Putting this together, the above expression simplifies to

\begin{equation}
    \begin{aligned}
    \label{eq:Td_Ch}
    (\td(X)\ch(Q))_4&= \frac{\rk(Q)}{1440} \left(60 \ch_3(X)\ch_1(X)+20 \ch_2(X)^2 - 168 \ch_4(X)\right)\\
     &\eqspace -\frac{1}{12}\ch_2(X)\ch_2(Q)+ \ch_4(Q)\,.
     \end{aligned}
\end{equation}

\subsection{Special case: \texorpdfstring{$Q= T^{1,0} \oplus T^{*1,0}$}{Q = T10 + T*01}}
Let us again consider the special case where $X$ is Calabi--Yau and $Q=T^{1,0} \oplus T^{*1,0}$ is the trivial direct sum of bundles. In this case\footnote{The observant reader may worry that even though the bundle $Q$ splits for the background geometry, the variation of the metric $\delta h$ will not respect this. However, it turns out that the offending deformations of $\delta h$ appear only off-diagonally in $h^{-1}\delta h$, and thus do not contribute when we take the trace.}
\begin{equation}
    \ch_n(Q)=2\ch_n(X)\,
\end{equation}
for $n$ even, while it vanishes for $n$ odd. We also have that $\rk(Q)=6$.

To analyse the metric anomaly, we vary the curvature as $\delta{\cal R}=\tfrac{1}{2\pi}g^{-1}\delta g=\tfrac{\gamma}{2\pi}$. We may write $\gamma$ as
\begin{equation}        
\gamma=\tfrac13\tr(\gamma)\id+\gamma_1=\gamma_0+\gamma_1\,,
\end{equation}
where $\gamma_1$ is trace-less and thus becomes an element in the adjoint of $SU(3)$. Recalling the expression for the partition function,
\begin{equation}
     |Z_{\text{free}}|^2 = \frac{I(Q)}{I(\Lambda^{0,0})^2}\, ,
\end{equation}
we see that a variation of the free energy with respect to the background metric $g$ can be written as
\begin{equation}
\delta\log|Z|=\frac{\pi}{2}\int_X\delta P_4({\cal R})\,,
\end{equation}
where the fourth-order curvature polynomial is
\begin{equation}
\begin{aligned}
 P_4({\cal R})&=\frac{\rk(Q)-2}{1440} \left(60 \ch_3(X)\ch_1(X)+20 \ch_2(X)^2 - 168 \ch_4(X)\right) \\
    &\eqspace -\frac{1}{12}\ch_2(X)\ch_2(Q)+ \ch_4(Q)\,,
    \end{aligned}
\end{equation}
and the $-2$ in the prefactor comes from the contribution of $I(\Lambda^{0,0})$ in the partition function. Consider first the metric variation due to $\gamma_0$. One finds\footnote{The reader might object to the slightly strange pre-factor in the expression. This factor becomes a bit more reasonable when gauge fields are included.}
\begin{equation}
\label{eq:CYMetricAnomaly0}
    \delta_0 \log|Z| = \frac{61}{360}\int_X \ch_3(X)\tr(\gamma)\,.
\end{equation}
Using the $SU(3)$ trace identity given in Appendix \ref{app:trace_identities}, $\ch_4(X)=\tfrac{1}{12}\ch_2(X)^2$, one finds the trace-free variation of the free energy, due to $\gamma_1$, is
\begin{equation}
\label{eq:CYMetricAnomaly1}
    \delta_1 \log |Z|=\frac{1}{120}\int_X\ch_2(X)\tr({\cal R}\gamma_1)\,.
\end{equation}
The first of these, $\delta_0 \log |Z|$, contributes to the standard Euler-class anomaly present in type IIB string theory or Kodaira--Spencer gravity~\cite{Bershadsky:1993cx, Bershadsky:1993ta, Pestun:2005rp}. In particular, when $\delta g$ corresponds to an on-shell Ricci-flat deformation, $\tr(\gamma_0)=\tr(g^{-1}\delta g)$ is constant and \eqref{eq:CYMetricAnomaly0} is proportional to the Euler class of $X$, $\chi(X)$.

As in Chern--Simons theory~\cite{Witten:1988hf}, one can cancel this non-trivial dependence on the background metric by adding appropriate local counter-terms. However, it seems that one cannot do this without imposing certain topological constraints. As an example, consider imposing the constraint that $\ch_2(X)$ is trivial. On a $\partial\bar\partial$-manifold, we then have
\begin{equation}
\label{eq:ConstraintMetricAnomalyCY}
    \ch_2(X)=\partial\bar\partial\omega^2_{\text{CS}}\, ,
\end{equation}
for a \emph{global} Chern--Simons two-form $\omega^2_{\text{CS}}\in\Omega^{1,1}(X)$, which we may take to be gauge invariant. The metric anomaly \eqref{eq:CYMetricAnomaly1} due to $\gamma_1$ may then be cancelled by adding the following purely background-dependent local counter-term to the action:
\begin{equation}
\label{eq:CounterMetricAnomalyCY}
    S_{1, \text{ct}}=\frac{\pi\ii}{120}\int_X\ch_2(X)\wedge\omega^2_{\text{CS}}\,.
\end{equation}
This is equivalent to multiplying the (absolute value of) the one-loop partition function with a purely background-dependent factor. As we want the partition function to be well-defined when traversing closed loops in the space of background metrics, it is natural to require the integrand of \eqref{eq:CounterMetricAnomalyCY} to be globally defined. Hence the need for the topological constraint \eqref{eq:ConstraintMetricAnomalyCY}. The variation of \eqref{eq:CounterMetricAnomalyCY} is
\begin{align}
  \delta S_{1,\text{ct}}&=-\frac{1}{120}\int_X\tr\left(\bar\partial\partial_\nabla(\gamma)\wedge {\cal R}\right)\wedge\omega^2_{\text{CS}}\notag\\
  &=\frac{1}{120}\int_X\ch_2(X)\wedge \tr({\cal R}\gamma_1)\,,  
\end{align}
which exactly cancels the term \eqref{eq:CYMetricAnomaly1}. 

In addition, if the Euler number of $X$ vanishes, so that is $\ch_3(X)$ is exact, the anomalous transformation \eqref{eq:CYMetricAnomaly0} due to $\gamma_0$ may be cancelled in a similar fashion. Indeed, we then have
\begin{equation}
    \ch_3(X)=\partial\bar\partial\omega^4_{\text{CS}}\,,
\end{equation}
for some globally well-defined Chern--Simons four-form $\omega^4_{\text{CS}}\in\Omega^{2,2}(X)$. We can then add a counter-term
\begin{equation}
    S_{0,\text{ct}}=\frac{61 \pi \ii }{180}\int_X\omega^4_{\text{CS}}\wedge \ch_1(X)\,,
\end{equation}
whose variation with respect to the background metric will cancel against \eqref{eq:CYMetricAnomaly0}. Alternatively, for non-vanishing Euler number, this part of the metric anomaly may be cancelled by multiplying the partition function by an appropriate volume factor, as was done in \cite{Pestun:2005rp}. This however requires that we stick to on-shell Ricci-flat deformations of the metric, where $\tr(\gamma)$ is constant.

\subsection{Including gauge fields}
We now consider the metric anomaly when we turn on gauge fields in our theory. This also means that we should study the twisted differential with fluxes $\bar D$ on $Q$. This complicates the metric anomaly quite a bit. It however turns out that if one re-introduces $\alpha'$ in the operator $\bar D$, the flux terms become higher order in the metric anomaly as they both appear at one-loop and are of order $\alpha'$. We show this explicitly in Appendix \ref{app:CurvPolQ}. This means that their contribution to the variation of the Ray--Singer torsions, and hence the one-loop partition function, will be at next order in the $\alpha', \hbar$ expansion. We will therefore ignore these terms for the remainder of this paper. We will leave the case of non-trivial flux for future work. Modulo higher orders, to get the metric anomaly it is therefore sufficient take the theory after the field redefinitions of Section \ref{sec:het_KS_gauge_fields} to be the untwisted model of section \ref{sec:untwisted_model}
\begin{equation}
\label{eq:localAction}
   S= \int_X\left( x \wedge \delb \chi + \k\, 
\wedge \delb b\right)+\int_X{\rm tr} (\alpha\,\bar\partial_A\alpha)\wedge\Omega\,,
\end{equation}
where $A$ denotes the background gauge connection. Note also that as the operator $\bar D$ is nilpotent, the corresponding Lagrangian, even with fluxes, can  an always locally be put in this form, modulo a local gauge transformation $\eta\in\Omega^{0,0}(\End(Q))$ such that $\bar D=\eta^{-1}\circ\bar\partial\circ \eta$. 

The partition function then becomes
\begin{equation}
    |Z_\text{free}|^2=\frac{I\left(T^{1,0} \oplus  T^{*1,0}\right)I(\End(V))}{I(\Lambda^{0,0})^2}\,.
\end{equation}
Varying the metric on $T^{1,0} \oplus  T^{*1,0}$ still corresponds to a choice of background metric on the Calabi--Yau manifold, while a choice of hermitian metric on the fibres of the gauge bundle $V$ is equivalent to a choice of Chern connection $A = h^{-1}\del h$. Hence, the metric anomaly for the gauge bundle $V$ can be identified with a background anomaly associated to a choice in background gauge field $A$. We note that the presence of such an anomaly does not mean the theory is ill, but merely indicates some background dependence of $h$ for the quantum theory. Nonetheless, we find that we can cancel this anomaly in special cases by introducing counter-terms, which we now show.

Including gauge fields, the metric anomaly polynomial now reads
\begin{equation}\label{eq:AnomalyPolGauge}
    \begin{aligned}
        P_4({\cal R},{\cal F})&= \frac{4+\dim G}{1440} \left(60 \ch_3(X)\ch_1(X)+20 \ch_2(X)^2 - 168 \ch_4(X)\right) \\
    &\eqspace-\frac{1}{12}\ch_2(X) \ch_2(\End(V))+ \ch_4(\End(V))\,,
    \end{aligned}
\end{equation}
where ${\cal F}=\frac{\ii}{2\pi}F$ is given by the curvature of the Chern connection on the gauge bundle. As for the metric anomaly associated to $T^{1,0}\oplus  T^{*1,0}$, we find that we can cancel the anomaly due to variation of the metric on the gauge bundle given certain topological constraints. In particular, we can cancel the anomaly via a background-dependent local counter-term as above if the background gauge group satisfies \eqref{eq:chern_character_constraint}.
A full list of simple Lie groups satisfying this constraint is given in Appendix \ref{app:trace_identities}, along with useful other trace identities. 

Furthermore, as for the variation of the Calabi--Yau metric, we can also include a singlet in the variation of the hermitian metric $\delta h$. However, this variation affects only the last two terms in \eqref{eq:AnomalyPolGauge}, and it drops out completely of the anomaly if the gauge bundle is self-dual. For more exotic gauge bundles, one can also imagine more exotic representations occurring in $\delta h$ (other than the singlet and the adjoint). This was not possible for the Calabi--Yau metric $g$ as in that case
\begin{equation}
    \delta g\in \boldsymbol{3}\times\boldsymbol{\bar 3}=\boldsymbol{1}+\boldsymbol{8}\,.
\end{equation}
However, if we again restrict to on-shell deformations, in that they preserve the Yang--Mills constraint
\begin{equation}
    \omega\wedge\omega\wedge F=0\,,
\end{equation}
we find that such more exotic representations must correspond to holomorphic sections of the given exotic representation of the bundle. Under suitable stability assumptions, there are no such sections. We will not consider these deformations for the remainder of the paper, and from now on take $h^{-1}\delta h$ to be adjoint valued. Let us now consider an example.

\subsubsection*{$G=SO(8)\times SO(8)$}
For $SO(8)$, the standard normalisation of the trace in the adjoint and fundamental representations implies that the Chern characters of $\End(V)$ and $V$ are related by
\begin{equation}
  \ch_2(\End(V))=6\ch_2(V)\,,\qquad
    \ch_4(\End(V))=\frac12 \ch_2(V)^2 \,.
\end{equation}
However, as we saw above, one copy of $SO(8)$ is not enough for the metric anomaly polynomial to factorise. This then makes it difficult to cancel the metric anomaly unless the second Chern character of both the tangent bundle and gauge bundle vanish separately.

If we instead use two $SO(8)$ groups, the anomaly polynomial reads
\begin{equation}\label{eq:Anomaly2SO(8)}
    \begin{aligned}
        P_4({\cal R},{\cal F}_1,{\cal F}_2)&= %
        \frac{19}{9}\ch_3(X)\ch_1(X)+\frac18\bigl(\ch_2(X)-2\ch_{2}(V_{1})\bigr)^2\\
        &\eqspace +\frac18\bigl(\ch_2(X)-2\ch_{2}(V_{2}))\bigr)^2\,,
    \end{aligned}
\end{equation}
where subscripts $1$ and $2$ refer to the two $SO(8)$ bundles. We note again that, although they decouple from the heterotic moduli problem, in order for the factorisation to work it was crucial to include the off-shell $(0,2)$-field $b$ and the $(3,0)$-field $\kappa$, related to a component of the Kalb-Ramond field and the axio-dilaton respectively. The anomaly can then be cancelled if both $SO(8)$ bundles have a second Chern character equalling half that of the tangent bundle of $X$. Note also then that the second Chern character of $V_1\oplus V_2$ must equal that of the tangent bundle, which agrees with the triviality of the heterotic Bianchi identity in ten dimensions. If we further impose the constraint coming from the flux twisted differential $\bar D$, i.e.~that the bundle $V_1\oplus V_2$ has vanishing second Chern class, we return to requiring that the second Chern class of $X$ vanishes. This is however not the case if we consider the untwisted theory of section \ref{sec:untwisted_model} instead.

One can consider other choices of product groups $G$. One finds that, modulo the term proportional to the Euler class of $X$, the polynomials take a form similar to those given in Appendix \ref{app:AnomalyPolynomials}. However, of all the combinations of gauge groups considered there, it is only $SO(8)\times SO(8)$ and $SU(3)\times E_6$ that automatically satisfy the ten-dimensional constraint that the second Chern class of the manifold equals the second Chern class of the gauge bundle, after imposing only the metric anomaly constraints on each individual bundle.

\section{Relation to holomorphic field theories}
\label{sec:Torsion}

The expansion of the heterotic superpotential around a Calabi--Yau background provides an example of a holomorphic field theory \cite{Williams:2018ows}. As we have shown, the one-loop partition function of this heterotic Kodaira--Spencer theory can be written as a product of holomorphic Ray--Singer torsions, giving a convenient way of analysing the anomaly structure in the case that the flux vanishes. We now discuss how to extend this to generic holomorphic field theories.
We will employ the path-integral formalism which is commonly found in the physics literature (and the previous sections of this article), rather than the precise mathematical constructions of~\cite{Williams:2018ows}, leaving a more rigorous construction for future study.

\subsection{One-loop quantisation in the BV formalism}

General Lagrangian field theories have a (background-dependent) perturbative description in terms of $L_\infty$ algebras (see~\cite{Hohm:2017pnh,Jurco:2018sby}). 
This result follows from the deformation theory of the equations of motion, which necessarily give an $L_{\infty}$ structure~\cite{Kontsevich-notes,Lada:1994mn,schlessinger2012deformation}. 
These algebras are characterised by a $\mathbb{Z}$-graded vector space $\mathcal{A}$ with a collection of $n$-fold graded-symmetric brackets $l_{n}\colon\mathcal{A}^{\otimes n} \to \mathcal{A}$, with intrinsic degree $2-n$, which satisfy Jacobi-like identities (see e.g.~\cite{Hohm:2017pnh} for details). In particular, the identities for the $l_{1} \equiv \dd$ bracket state that
\begin{equation}
    \dd^{2} = 0\, ,
\end{equation}
and hence $\mathcal{A}$ forms a differential complex:
\begin{equation}
\label{eq:gauge-complex}
	\mathcal{A} \colon \quad \ldots \stackrel{\dd}{\longrightarrow} A^{0} \stackrel{\dd}{\longrightarrow} A^1 \stackrel{\dd}{\longrightarrow} A^{2} \stackrel{\dd}{\longrightarrow} \ldots
\end{equation}
Note that $\dd$ may not be a first-order differential operator in general. The $\mathbb{Z}$-grading coincides with the BRST grading of the underlying theory. For example, the fields live in $A^{1}$, the gauge parameters live in $A^{0}$, and so on. The equations of motion live in $A^{2}$ and can be written as
\begin{equation}
    0 = \dd y + l_{2}(y,y) + l_{3}(y,y,y) +\ldots \qquad \text{with }y \in A^{1}\, .
\end{equation}

Working at the one-loop level, we can restrict to the linearised theory. We can therefore drop all of the higher brackets and consider simply the differential complex \eqref{eq:gauge-complex}. For Lagrangian theories, one can find a natural pairing $\langle \cdot,\cdot \rangle $ on $\mathcal{A}$ such that the classical action can be expressed as
\begin{equation}
\label{eq:action}
	S = \langle y , \dd y \rangle \sim \int  y \,\dd y \qquad \text{with }y \in A^{1}\, .
\end{equation}
In general, this pairing is required to satisfy identities which correspond to ``invariance'' under $L_\infty$ bracket operations (usually referred to as a cyclic structure). 
In the linearised case, this simply translates into integration by parts. 
When the equations of motion are those solved by a deformation of a solution to some fixed theory, as was the case for our heterotic superpotential theory, one can derive this pairing from the original action.

Introducing ghosts, antifields and so on, as in the BV formalism, we find that they lie in the remaining vector spaces $A^{n}$. The minimal solution to the master action can then be written in the same form but now with $A$ a generic element of \eqref{eq:gauge-complex}~\cite{Jurco:2018sby}:
\begin{equation}
\label{eq:master-action}
	S = \langle y , \dd y \rangle \sim \int  y \, \dd y
	\qquad \text{with } y \in \mathcal{A}\, .
\end{equation}
To quantise the theory, one introduces a Hodge star-like operator $*$ such that the new inner product
\begin{equation}
\label{eq:inner-prod}
	(y,y) = \langle y , \ast y \rangle
\end{equation}
is positive definite.\footnote{In the case of complex fields, this star operator includes complex conjugation so as to ensure positivity.} Using this, one can define the adjoint of $\dd$ such that the resulting Laplacian is elliptic and choose a gauge-fixing condition $\dd^\dagger y = 0$.  
The modulus of the one-loop partition function of the theory defined by \eqref{eq:master-action} is then given in terms of the analytic torsion of the differential complex~\eqref{eq:gauge-complex}~\cite{Schwarz:1978cn,Schwarz:1979ae}. 
For simplicity, let us assume that the field $A^1$ is bosonic and that the gauge symmetry is finitely reducible. 
We can then express the partition function concisely as:
\begin{equation}
\label{eq:Z-one-loop}
	|Z_{\mathcal{A}}|^2 = \AT^{(-1)^{(N+2)/2}}\, ,
\end{equation}
where $N$ is the number of non-trivial terms in the complex $\mathcal{A}$ and 
\begin{equation}
\label{eq:analytic-torsion-general}
	\AT = \Bigl( \prod_{p} (\det \Delta^{p})^{(-1)^{p}p} \Bigr)^{1/2}
\end{equation}
is the analytic torsion of $\mathcal{A}$. Note that in~\eqref{eq:analytic-torsion-general} we have introduced a shifted grading $p$ on the complex $\mathcal{A}$ defined such that the lowest-degree vector space in $\mathcal{A}$ now has label $p=0$.

\subsection{Analytic torsion and Dolbeault resolution}
\label{sec:Dolbeault-torsion}

Holomorphic field theories are a particular type of field theory whose BRST complex $\mathcal{A}$ is in fact the total complex of a double complex of a certain type. This double complex is, in particular, the Dolbeault resolution of a single holomorphic complex. That is, we have some complex $\mathcal{C}^{\bullet}$ which consists of holomorphic sections of some vector bundles $C^{\bullet}$, with a holomorphic differential operator $\Qhol$ which squares to zero:
\begin{equation}\label{eq:sheaf_complex}
    0 \stackrel{}{\longrightarrow} \mathcal{C}^{0} \stackrel{\Qhol}{\longrightarrow} \mathcal{C}^{1} \stackrel{\Qhol}{\longrightarrow}  \mathcal{C}^{2} \stackrel{\Qhol}{\longrightarrow} \ldots
\end{equation}
A natural example of such a $\Qhol$ is the Dolbeault differential $\del$ -- this is the holomorphic differential appearing in~\eqref{eq:L_inft_redef}, which thus underlies the heterotic superpotential theory we have been studying. 

Sheaves of holomorphic sections always have a Dolbeault resolution, and hence in general we can picture the double complex as
\begin{equation}
\label{eq:holFT-complex}
\begin{tikzpicture}[scale=1.5,baseline=(current bounding box.center)] %
\node (z) at (-2.5,2) {}; 
\node (h) at (-1.3,2) {$0$}; 
\node (0) at (0,2) {$0$}; 
\node (1) at (1.5,2) {$0$}; 
\node (2) at (3.2,2) {$0$}; 
\node (3) at (4.5,2) {}; 
\node (Az) at (-2.5,1) {$0$}; 
\node (Ah) at (-1.3,1) {$\mathcal{C}^{0}$}; 
\node (A0) at (0,1) {$\Omega^{0,0}(C^{0})$}; 
\node (A1) at (1.5,1) {$\Omega^{0,1}(C^{0})$}; 
\node (A2) at (3.2,1) {$\Omega^{0,2}(C^{0})$}; 
\node (A3) at (4.5,1) {}; 
\node (Bz) at (-2.5,0) {$0$}; 
\node (Bh) at (-1.3,0) {$\mathcal{C}^1$}; 
\node (B0) at (0,0) {$\Omega^{0,0} (C^1)$}; 
\node (B1) at (1.5,0) {$\Omega^{0,1}(C^1)$}; 
\node (B2) at (3.2,0) {$\Omega^{0,2}(C^1)$}; 
\node (B3) at (4.5,0) {}; 
\node (Cz) at (-2.5,-1) {$0$}; 
\node (Ch) at (-1.3,-1) {$\mathcal{C}^2$}; 
\node (C0) at (0,-1) {$\Omega^{0,0} (C^2)$}; 
\node (C1) at (1.5,-1) {$\Omega^{0,1}(C^2)$}; 
\node (C2) at (3.2,-1) {$\Omega^{0,2}(C^2)$}; 
\node (C3) at (4.5,-1) {}; 
\node (Dz) at (-2.5,-2) {}; 
\node (Dh) at (-1.3,-2) {}; 
\node (D0) at (0,-2) {}; 
\node (D1) at (1.5,-2) {}; 
\node (D2) at (3.2,-2) {}; 
\node (D3) at (4.5,-2) {}; 
\path[->,font=\scriptsize] 
(Az) edge node[above]{} (Ah)
(Ah) edge node[above]{$\iota$} (A0)
(A0) edge node[above]{$\bar\der$} (A1)
(A1) edge node[above]{$\bar\der$} (A2)
(A2) edge node[above]{$\bar\der$} (A3)
(Bz) edge node[above]{} (Bh)
(Bh) edge node[above]{$\iota$} (B0)
(B0) edge node[above]{$\bar\der$} (B1)
(B1) edge node[above]{$\bar\der$} (B2)
(B2) edge node[above]{$\bar\der$} (B3)
(Cz) edge node[above]{} (Ch)
(Ch) edge node[above]{$\iota$} (C0)
(C0) edge node[above]{$\bar\der$} (C1)
(C1) edge node[above]{$\bar\der$} (C2)
(C2) edge node[above]{$\bar\der$} (C3)
(h) edge node[left]{} (Ah)
(0) edge node[left]{} (A0)
(1) edge node[left]{} (A1)
(2) edge node[left]{} (A2)
(Ah) edge node[left]{$\Qhol$} (Bh)
(A0) edge node[left]{$\Qhol$} (B0)
(A1) edge node[left]{$\Qhol$} (B1)
(A2) edge node[left]{$\Qhol$} (B2)
(Bh) edge node[left]{$\Qhol$} (Ch)
(B0) edge node[left]{$\Qhol$} (C0)
(B1) edge node[left]{$\Qhol$} (C1)
(B2) edge node[left]{$\Qhol$} (C2)
(Ch) edge node[left]{$\Qhol$} (Dh)
(C0) edge node[left]{$\Qhol$} (D0)
(C1) edge node[left]{$\Qhol$} (D1)
(C2) edge node[left]{$\Qhol$} (D2);
\end{tikzpicture}
\end{equation}
The left-most column here consists of the sheaves of holomorphic sections in \eqref{eq:sheaf_complex} that are being resolved.

Writing $B^{p,q} = \Omega^{0,p}(C^{q})$, we say that our theory is a \emph{holomorphic} theory if the BRST complex $\mathcal{A}$ decomposes into a double complex as
\begin{equation}
    A^{n} = \bigoplus_{p+q=n}B^{p,q} = \bigoplus_{p+q=n} \Omega^{0,p}(C^{q})\, .
\end{equation}
The claim is that under certain circumstances, the analytic torsion of $\mathcal{A}$, and hence the one-loop partition function of the theory, can then be expressed in terms of the holomorphic torsions\footnote{To be clear, we write analytic torsion for an arbitrary comblex $(\mathcal{A},\dd)$, while the holomorphic torsion is reserved for the analytic torsion of the Dolbeault complex of a holomorphic vector bundle $(\Omega^{0,\bullet}(C),\delb)$.} of the holomorphic bundles $C^q$.

In order to define Laplacians, we introduce a positive-definite inner product on \eqref{eq:holFT-complex}, similar to that in~\eqref{eq:het+ve-prod}. Writing the differential on the total complex as $\dd_T = \bar\der + \Qhol$, we use this inner product to define its adjoint $\dd_T^\dagger$ and a Laplacian $\Delta_T = \dd_T \dd_T^\dagger + \dd_T^\dagger \dd_T$. 
In the case that the Laplacian preserves the individual spaces $B^{p,q}$, which is the case for $\Qhol=\del$ on a K\"ahler manifold, let $\Delta^{p,q}_T$ denote the Laplacian acting on $B^{p,q}$.
The analytic torsion of $\mathcal{A}$ can then be written as\footnote{Note that the exponent here is different to the definition in~\cite{Ashmore:2021pdm} due to a difference in the grading of the total complex.}
\begin{equation}
\label{eq:total-torsion}
	\AT = \Bigl( \prod_{p,q} (\det \Delta_T^{p,q})^{(-1)^{p+q}(p+q)} \Bigr)^{1/2}\, .
\end{equation}
The total analytic torsion of the Dolbeault resolution of the bundle $C^{q}$ is
\begin{equation}
\label{eq:pth-torsion}
	T_{q} = \Bigl( \prod_{p} (\det \Delta_T^{p,q})^{p\, (-1)^{p}} \Bigr)^{1/2}\, ,
\end{equation}
and we define the total analytic torsion of the $\Omega^{0,p}$-twisted complex $(\Omega^{0,p}(C^{\bullet}),\Qhol)$ to be
\begin{equation}
\label{eq:Sp}
	S_p = \Bigl( \prod_{q} (\det \Delta_T^{p,q})^{q(-1)^{q}} \Bigr)^{1/2}\, .
\end{equation}
It is then simple to check that the total torsion factorises into products of torsions $T_{q}$ and $S_{p}$ as follows:
\begin{equation}
\label{eq:TT-formula}
	   \AT = \Bigl( \prod_{q} T_q^{(-1)^q} \Bigr)
		\Bigl( \prod_{p} S_{p}^{(-1)^{p}} \Bigr)\, .
\end{equation}
We observe that if $\Delta_{T}$ commutes with $\delb$ and $\delb^{\dagger}$, then it preserves the decomposition\footnote{In considering the one-loop partition function, we always remove the zero-modes.}
\begin{equation}\label{eq:decomposition}
    B^{p,q} = \delb B^{p-1,q} \oplus \delb^{\dagger} B^{p+1,q}\, .
\end{equation}
In particular, this holds if $\delb$ and $\Qhol$ satisfy certain K\"ahler-like identities. If this is the case, then, in line with \cite{Pestun:2005rp,Ashmore:2021pdm}, we can define operators $\smallbullet\Delta_{T}$ and $\Delta_{T}\smallbullet$ which are the restrictions of $\Delta_{T}$ to the first and second subspaces respectively in the decomposition \eqref{eq:decomposition}. Moreover, since we have
\begin{equation}
\begin{aligned}
    \Delta^{p,q}_{T} &= \smallbullet \Delta^{p,q}_{T} + \Delta^{p,q}_{T}\smallbullet \, ,\\
    \smallbullet\Delta_{T}^{p,q}\delb \alpha &= \Delta^{p,q}_{T}\delb \alpha = \delb \Delta_{T}^{p-1,q}\alpha = \delb(\Delta_{T}^{p-1,q}\smallbullet \alpha)\, ,
\end{aligned}
\end{equation}
the determinant of the total Laplacian can be expressed as
\begin{equation}
    \det \Delta_{T}^{p,q} = (\det \smallbullet\Delta_{T}^{p,q}) (\det \Delta_{T}^{p,q}\smallbullet) \, ,\qquad \det\smallbullet\Delta^{p,q}_{T} = \det \Delta^{p-1,q}_{T}\smallbullet\, .
\end{equation}
It is then a quick calculation to check
\begin{equation}
    \prod_{p}S_{p}^{(-1)^{p}} = 1\, .
\end{equation}
We therefore find that the total torsion of $\mathcal{A}$, i.e.~the one-loop partition function, is the alternating product of the total analytic torsions of the Dolbeault resolutions of each of the $C^{q}$. That is, we have
\begin{equation}
\label{eq:Z-hol-torsion}
    \AT = \prod_{q} T_q^{(-1)^q} \, .
\end{equation}
We emphasise, though, that in this formula the torsions $T_q$ are those defined in~\eqref{eq:pth-torsion}, featuring the determinants of the Laplacian $\Delta_T$ for the total complex, and not those with determinants of the standard Dolbeault Laplacians $\Delta_{\bar\der}$ that would give the holomorphic Ray--Singer torsions $I(C^q)$. 
It would thus appear that in order to relate the quantity~\eqref{eq:pth-torsion} to these holomorphic Ray--Singer torsions we need to assume a further direct relation between $\det \Delta^{p,q}_T$ and $\det \Delta^{p,q}_{\bar\der}$, as we found for the heterotic superpotential in Equation~\eqref{eq:det-equality}. 
However, as we will see below, remarkably, %
the overall expression for $\AT$ is equal to that with $I(C^q)$ in place of $T_q$ provided some considerably milder assumptions hold.

\subsection{Alternative approach to the holomorphic torsion}
\label{sec:hol-field-redef}

We will now show 
that the analytic torsion~\eqref{eq:Z-hol-torsion} is equal to the alternating product of holomorphic Ray--Singer torsions $I(C^q)$ for certain free holomorphic field theories~\cite{Williams:2018ows} (specified below) by 
implementing a field redefinition which generalises that used in 
Section~\ref{sec:het_KS_gauge_fields}. 

Consider a free holomorphic field theory with BV complex~\eqref{eq:holFT-complex} (minus the left-most column). In this subsection, we will need the following assumptions which restrict both the choice of theory and the underlying complex manifold. First, we assume that 
\begin{equation}
   [\bar\der, \Qhol] = 0 
    \qquad \text{and} \qquad 
   [\bar\der^\dagger, \Qhol] = 0  \,.
\end{equation}
This would be true, for example, in the case where $\Qhol$ is a holomorphic Dolbeault operator acting on differential forms and the manifold is K\"ahler. We also assume that we have a Hodge decomposition for the operator $\bar\der$, which is guaranteed if the manifold is compact. 
We will consider the theory absent of zero-modes, though this could be weakened simply to 
assuming that $\Qhol$ annihilates the zero-modes of $\Delta_{\bar\der}$. 

Denoting a general element of the BV complex by $y$ and the inner product as $\langle\cdot ,\cdot\rangle$, the BV master action takes the same form as above,
\begin{equation}
	S = \langle y , \dd y \rangle = \langle y , \bar\der y \rangle + \langle y , \Qhol y \rangle \, , %
\end{equation}
where $\dd = \bar\der + \Qhol$ is the natural derivative on the total complex of~\eqref{eq:holFT-complex}. Let us now apply the Hodge decomposition for $\bar\der$ to the field $y$ by writing
\begin{equation}
   y = \bar\der b + \bar\der^\dagger c \, . %
\end{equation}
Note that $b$ and $c$ can also be thought of as elements of the double complex. 

Via (anti-)commutation relations of operators and integration by parts, the action then becomes, schematically (up to irrelevant signs which depend on the $(p,q)$ degree of $y$), 
\begin{equation}
\begin{aligned}
	S &= \langle y , \bar\der y \rangle 
		+ \langle y , \Qhol \bar\der b \rangle
		+ \langle y , \Qhol \bar\der^\dagger c \rangle \\
		&= \langle y , \bar\der y \rangle 
		+ \langle y , \Qhol \bar\der b \rangle
		+ \langle  \bar\der b ,   \Qhol \bar\der^\dagger c \rangle \\
		&= \langle y , \bar\der y \rangle 
		+ \langle y , \Qhol \bar\der b \rangle
		+ \langle  \bar\der b ,   \Qhol y \rangle \\
		&= \langle y , \bar\der y \rangle 
		+ 2 \langle y , \Qhol \bar\der b \rangle \\
		&= \langle y + \Qhol b , \bar\der (y + \Qhol b) \rangle \\
		& = \langle y' , \bar\der y' \rangle 
			\qquad \text{where } y' = y + \Qhol b\, . \\
\end{aligned}
\end{equation}
For each component of definite bi-degree $(p,q)$ of the field $y$, the corresponding component of the field $b$ has bi-degree $(p,q-1)$, while $\Qhol b$ has bi-degree $(p+1,q-1)$. Therefore, in the shift $y' = y + \Qhol b$ each component of $y$ of bi-degree $(p,q)$ is shifted by a term depending only on the component of $y$ of bi-degree $(p-1,q+1)$. The shift of the total field $y$ is thus upper triangular if decomposed with respect to an overall $q-p$ grading. Therefore, the (non-local) field redefinition from $y$ to $y'$ will have trivial Jacobian in the path integral.

This result implies that the free holomorphic field theory is equivalent to one with the same field content but $\Qhol = 0$, even at one loop. In particular, this should be true for any free holomorphic field theory on a compact K\"ahler $n$-fold where the fields live in 
vector bundles associated to the $U(n)$ structure principal bundle and the operator $\Qhol$ is constructed from Levi-Civita connections and $U(n)$ invariant tensors.

An immediate corollary of this statement is that the analytic torsion $\AT$ (and thus the modulus of the partition function by~\eqref{eq:Z-one-loop}) is given by the alternating product of holomorphic Ray--Singer torsions of the holomorphic bundles~$C^p$, as in Equation~\eqref{eq:Z-hol-torsion}, but with $T_p$ replaced by $I(C^p)$:
\begin{equation}
\label{eq:Z-hol-torsion2}
	|Z_{\mathcal{A}}|^2 = \AT^{(-1)^{(N+2)/2}}\, ,
	\quad
	\text{where }
    \AT =  \prod_{q} I(C^q)^{(-1)^q}\, .
\end{equation}
As claimed, we have arrived at this result utilising only the assumptions detailed at the start of this subsection, which are considerably weaker than those needed in Section \ref{sec:Dolbeault-torsion}.

\subsection{Application to heterotic superpotential}

The BV gauge-complex for the heterotic superpotential theory is:
\begin{equation}
\label{eq:het-BV-complex}
\begin{tikzpicture}[scale=1.5,baseline=(current bounding box.center)] %
\node (z) at (-2.5,2) {}; 
\node (h) at (-1.3,2) {0}; 
\node (0) at (0,2) {0}; 
\node (1) at (1.6,2) {0}; 
\node (2) at (3.2,2) {0}; 
\node (3) at (4.8,2) {0}; 
\node (Az) at (-2.5,1) {$0$}; 
\node (Ah) at (-1.3,1) {$\mathcal{O}$}; 
\node (A0) at (0,1) {$\Omega^{0,0}$}; 
\node (A1) at (1.6,1) {$\Omega^{0,1}$}; 
\node (A2) at (3.2,1) {$\Omega^{0,2}$}; 
\node (A3) at (4.8,1) {$\Omega^{0,3}$}; 
\node (A4) at (6.1,1) {$0$}; 
\node (Bz) at (-2.5,0) {$0$}; 
\node (Bh) at (-1.3,0) {$\mathcal{Q}$}; 
\node (B0) at (0,0) {$\Omega^{0,0} (Q)$}; 
\node (B1) at (1.6,0) {$\Omega^{0,1}(Q)$}; 
\node (B2) at (3.2,0) {$\Omega^{0,2}(Q)$}; 
\node (B3) at (4.8,0) {$\Omega^{0,3}(Q)$}; 
\node (B4) at (6.1,0) {$0$}; 
\node (Cz) at (-2.5,-1) {$0$}; 
\node (Ch) at (-1.3,-1) {$\mathcal{O}$}; 
\node (C0) at (0,-1) {$\Omega^{3,0} $}; 
\node (C1) at (1.6,-1) {$\Omega^{3,1}$}; 
\node (C2) at (3.2,-1) {$\Omega^{3,2}$}; 
\node (C3) at (4.8,-1) {$\Omega^{3,3}$}; 
\node (C4) at (6.1,-1) {$0$}; 
\node (Dz) at (-2.5,-2) {}; 
\node (Dh) at (-1.3,-2) {0}; 
\node (D0) at (0,-2) {0}; 
\node (D1) at (1.6,-2) {0}; 
\node (D2) at (3.2,-2) {0}; 
\node (D3) at (4.8,-2) {0}; 
\path[->,font=\scriptsize] 
(Az) edge node[above]{} (Ah)
(Ah) edge node[above]{$\iota$} (A0)
(A0) edge node[above]{$\bar\der$} (A1)
(A1) edge node[above]{$\bar\der$} (A2)
(A2) edge node[above]{$\bar\der$} (A3)
(A3) edge node[above]{} (A4)
(Bz) edge node[above]{} (Bh)
(Bh) edge node[above]{$\iota$} (B0)
(B0) edge node[above]{$\bar{D}$} (B1)
(B1) edge node[above]{$\bar{D}$} (B2)
(B2) edge node[above]{$\bar{D}$} (B3)
(B3) edge node[above]{} (B4)
(Cz) edge node[above]{} (Ch)
(Ch) edge node[above]{$\iota$} (C0)
(C0) edge node[above]{$\bar\der$} (C1)
(C1) edge node[above]{$\bar\der$} (C2)
(C2) edge node[above]{$\bar\der$} (C3)
(C3) edge node[above]{} (C4)
(h) edge node[left]{} (Ah)
(0) edge node[left]{} (A0)
(1) edge node[left]{} (A1)
(2) edge node[left]{} (A2)
(3) edge node[left]{} (A3)
(Ah) edge node[left]{$\der$} (Bh)
(A0) edge node[left]{$\der$} (B0)
(A1) edge node[left]{$\der$} (B1)
(A2) edge node[left]{$\der$} (B2)
(A3) edge node[left]{$\der$} (B3)
(Bh) edge node[left]{$\der$} (Ch)
(B0) edge node[left]{$\der$} (C0)
(B1) edge node[left]{$\der$} (C1)
(B2) edge node[left]{$\der$} (C2)
(B3) edge node[left]{$\der$} (C3)
(Ch) edge node[left]{} (Dh)
(C0) edge node[left]{} (D0)
(C1) edge node[left]{} (D1)
(C2) edge node[left]{} (D2)
(C3) edge node[left]{} (D3);
\end{tikzpicture}
\end{equation}
When we perturb around a background where the fluxes vanish, the operator $\bar{D} \equiv \delb$ and the system has precisely the form outlined in the previous subsections. We also have the stronger condition that the determinant of the total Laplacian is proportional to that of the Dolbeault Laplacian, which implies $\det\Delta_{T}^{p,q} \propto \det\Delta_{\delb}^{p,q}$. In particular, this implies that the total analytic torsion of the Dolbeault resolution of $C^{p}$ is the holomorphic Ray--Singer torsion of $C^{p}$:
\begin{equation}
    T_{p} = I(C^{p})\, .
\end{equation}
Therefore, either by using this together with~\eqref{eq:Z-hol-torsion} or by directly applying the more general formula~\eqref{eq:Z-hol-torsion2}, we arrive at the expected result:
\begin{equation}
    |Z_\text{free}|^2 =  \prod_{p} T_p^{(-1)^p} = \frac{I(Q)}{I(\Lambda^{0,0})I(\Lambda^{3,0})} = \frac{I(Q)}{I(\Lambda^{0,0})^{2}}\, .
\end{equation}

We find it suggestive that, provided the BRST complex satisfies certain K\"ahler-like identities, the one-loop partition function of a holomorphic field theory can be written as a product of holomorphic Ray--Singer torsions, and that these conditions precisely arise for the heterotic superpotential theory. %
This means that the anomalies of such field theories, both background and gravitational/gauge anomalies, can be analysed very easily using the same methods we have used in this paper.

\section{Conclusions}\label{sec:conclusions}

In this work, we have studied a topological theory whose equations of motion give the Maurer--Cartan equation associated to the Hull--Strominger system at $\mathcal{O}(\alpha')$ \cite{Ashmore:2018ybe}. It effectively describes Kodaira--Spencer theory coupled to hermitian and gauge bundle deformations, and hence can be considered as the heterotic extension of Kodaira--Spencer theory \cite{Bershadsky:1993cx}. Our work extends that of \cite{Ashmore:2018ybe} by including rescalings of the holomorphic three-form $\Omega$, or deformations associated to the dilaton.

Working at quadratic order, we were able to find the one-loop partition function in terms of holomorphic Ray--Singer torsions of holomorphic bundles \cite{bismut1988analytic1,bismut1988analytic2,bismut1988analytic3}. We analysed the anomalies of the background to first order in $\alpha'$, and were able to identify two different kinds -- background metric anomalies and gravitational/gauge anomalies. Background anomalies are associated to a change of the background geometry one is perturbing. Their presence does not signify an inconsistent theory, just that the quantum theory depends on more data than one expected. Nonetheless, provided the second Chern class of the manifold and the gauge bundle are correlated in a way depending on the gauge group, we found that we could cancel the dependence on a choice of hermitian bundle metrics, and hence the theory retains its topological nature at one-loop. %

The gauge/gravitational anomalies on the other hand are associated to a change in local frame and they do signify an inconsistent theory. We found that only gravitational anomalies could arise and that these can be cancelled through a Green--Schwarz type mechanism. We also defined a slightly different holomorphic field theory with somewhat weaker constraints on the construction. In doing so, we found that this model had gauge, gravitational and mixed anomalies. Again, these can be cancelled assuming that the Chern characters of the gauge bundle and background are correlated, as in the background anomaly calculation. Particularly interesting were the cases where the gauge group was $SO(8)\times SO(8)$ or $SU(3)\times E_6$, where the anomaly cancellation constraint coincided with that of the ten-dimensional heterotic string. The gauge group $SU(3)\times E_6$ is interesting as it corresponds to the internal gauge group of the standard embedding~\cite{Candelas:1985en}.

In the final part of the paper, we made some observations on the link between this heterotic Kodaira--Spencer theory and holomorphic field theories \cite{Williams:2018ows}. In particular, we noted that when the flux vanishes, the action defines a holomorphic theory, and we gave a novel derivation of the fact that the one-loop partition function should be given by the product of holomorphic Ray--Singer torsions.

Our work leads to many natural and interesting questions. Firstly, one should hope to be able to define the theory in other complex dimensions, as for the original Kodaira--Spencer theory~\cite{Costello:2012cy}. 
Indeed, there is a clear path to doing so on $\SU{N}$-structure manifolds in $2N$ dimensions as follows. 
Under $\SU{N}$, one can identify chiral spinors with even or odd anti-holomorphic forms. 
One can then easily see that the fermionic fields of heterotic supergravity (plus the corresponding antifields) fill out a complex similar to that of our theory given in~\eqref{eq:het-BV-complex}. 
One should thus expect that these will provide the higher-degree form fields needed to write a BV master action of Chern--Simons or BF type which will include the degrees of freedom for the deformations of the complex / K\"ahler structures and gauge fields at lower degrees. 
This is what happens in Kodaira--Spencer theory~\cite{Costello:2012cy} and also twisted super Yang--Mills in ten dimensions~\cite{Baulieu:2010ch} -- one should expect our case to go through similarly. 

Next, in the regime where the fluxes have been turned off, the bundle $Q=T^{1,0}\oplus \End(V)\oplus T^{*1,0}$ is a direct sum of bundles. When the fluxes are reintroduced, the bundle $Q$ takes this form locally, but globally has an interesting extension structure. Alternatively, one can view the differential $\bar{D}$ as no-longer being diagonal but instead upper triangular. 
Crucially though, $Q$ remains a holomorphic vector bundle. 
Thus, one can build the analogue of the Dolbeault resolution~\eqref{eq:het-BV-complex}, but now with the full anti-holomorphic differential $\bar{D}$ rather than $\bar\der$ for the terms involving $Q$. 
Because of this, it no longer strictly fits into the definition of a holomorphic field theory in~\cite{Williams:2018ows}, though it is clearly very similar in spirit. 
It can be shown that the construction of section~\ref{sec:hol-field-redef} otherwise goes through without modification in this setting, leaving the same form for the absolute value of the partition function, but now defining the holomorphic torsion of $Q$ using the operator $\bar{D}$. 
One could then go on to examine its anomalies, though this is likely to be less straightforward as some ingredients of the theorems providing the variations of the torsions would then be absent. 
One could hope, however, that the theory would remain well-defined with non-zero background flux. 

Further, including the fluxes also leads us to consider other anomalies of both background and gauge type. In particular, there may be anomalies associated to the choice of $B$-field, which can be combined with the metric anomaly to define a \emph{generalised metric anomaly}. The robust nature of anomalies lead to the identification of the conventional anomaly polynomial with the index of the Dirac operator on the manifold. In the case of the $B$-field, it does not define a conventional connection but instead a connective structure on a gerbe, and hence the natural anomaly may be in terms of indices for gerbe connections. A modern physical interpretation of the $B$-field is the gauge field for a $U(1)$ one-form symmetry. This theory may be an interesting arena to probe the link between higher-form gauge anomalies and gerbes. The authors expect that finding a covariant form of the anomaly-cancelling counter-terms in Section \ref{sec:metric_anomaly} that respect the structure of $Q$ would provide insight into this question.

In this paper, we restricted to studying the one-loop partition function of the theory, or equivalently the free theory. Including the higher-order interactions would then give a new interacting holomorphic field theory. One could try to find the associated $L_{\infty}$ algebra and the associated BV action, extending the work of \cite{Ashmore:2018ybe} to include the dilaton. Calculating higher loop partition functions including these interactions may lead to new invariants of the holomorphic Courant algebroids. This is similar to what happens for the topological A/B model and holomorphic Chern-Simons. In practice, however, this may prove to be difficult as the higher interaction contain holomorphic derivatives. Understanding the holomorphic anomaly in this context may allow us to make progress in calculating the higher loop partition functions.

It is possible to reinterpret the topological action in this paper in terms of the language of exceptional complex structures in $O(6,6+n)$ generalised geometry~\cite{Ashmore:2019rkx,Smith:2022baw}. 
In this picture, the tensors defining the Hull--Strominger system can combine to define a complex vector bundle $L$ within the (complexified) generalised tangent bundle $E_{\mathbb{C}}$. 
The F-terms, i.e.~the vanishing of $W$ and $\delta W$, are equivalent to involutivity of $L$ under the Dorfman derivative plus a additional mild condition related to the dilaton. The Maurer--Cartan equation given in \cite{Ashmore:2018ybe} is then precisely the Maurer--Cartan equation associated to deformations of $L$ which preserve involutivity. 
This makes the relationship to Kodaira--Spencer theory even more obvious. 
The exceptional complex structure has associated to it a generalised Dolbeault complex, whose second cohomology computes its infinitesimal moduli. 
In fact, this can be defined for any generalised $G$-structure, in perfect analogy to the usual Dolbeault complexes associated to torsion-free $G$-structures, as in~\cite{ReyesCarrion:1998si}. 
For generalised metric compatible $G$-structures, it becomes a double complex, as in~\cite{Ashmore:2021pdm}. 
A small extension of this double complex to account for the dilaton (which can be motivated by requiring a cyclic structure and hence an action), then provides a generalised geometric construction of the double complex~\eqref{eq:het-BV-complex}. 
Further details of this construction will be reported elsewhere. 

As was shown in \cite{Ashmore:2019qii,Tennyson:2021qwl}, analogous structures appear generically in M-theory and type II compactifications. We can repeat the analysis done here to define and study a Kodaira--Spencer theory coupled to RR degrees of freedom for type II, or a Kodaira--Spencer theory for $G_{2}$ backgrounds in M-theory, each derived from an $\mathcal{N}=1$ superpotential. These constructions will also be reported elsewhere.

Finally, it would be interesting to investigate the connection between our topological theory and a topological heterotic string. Given the link between Kodaira--Spencer and the topological B-model \cite{Bershadsky:1993cx},\footnote{In fact it was shown in \cite{Pestun:2005rp} that Kodaira--Spencer theory fails to match the B-model at one-loop. Instead, the B-model is more accurately described by a theory of generalised complex structures. \cite{Ashmore:2021pdm} found a similar result for the conjectural $G_{2}$ and $Spin(7)$ topological strings. The fact that our action is already related to deformations of exceptional complex structures lends greater evidence to our claim.} it is natural to suppose that our heterotic Kodaira--Spencer theory could be the target-space theory of such a topological heterotic string. Candidates for this topological string include analogues of Witten's open string~\cite{Witten:1992fb}, holomorphically twisted $(0,2)$-models~\cite{Katz:2004nn, Adams:2005tc, Sharpe:2006qd}, or analogues of the well-studied $\beta\gamma$-system~\cite{Witten:2005px, Kapustin:2005pt, Nekrasov:2005wg}. In fact, a recent paper studied a duality of Lie algebras of the chiral sector of the $\beta\gamma$-system and a ten-dimensional target space theory of similar flavour to ours~\cite{Costello:2021kiv}. 

A first statement in this direction could be a version of the conjecture of Costello and Li that the ``minimal'' BCOV theory is in fact a twisted version of type IIB supergravity on $\mathbb{C}^5$~\cite{Costello:2016mgj}, where the larger set of fields considered there provides a candidate for the twist of the full type IIB string field theory.
In our context, one could guess that the superpotential theory describes the six-dimensional part of a twisted version of heterotic supergravity on the background $\mathbb{R}^{4} \times \text{CY}_3$. 
Establishing this would provide a stepping stone towards grander statements concerning the full string theory. 

Related to this is the question of holomorphic anomalies. While we did not highlight it in this paper, the one-loop partition function suffers from a holomorphic anomaly which can be calculated following \cite{Bershadsky:1993cx,bismut1988analytic1,bismut1988analytic2,bismut1988analytic3}. In \cite{Bershadsky:1993cx}, such anomalies for the topological B-model were shown to arise due to boundaries in complex structure moduli space, and they used this to find a recursion relation for the partition function at different loop orders. One might ask whether our holomorphic anomaly arises similarly and can be used to determine higher-order partition functions of the topological heterotic string. Given that the topological A- and B-models played a great role in formulating and testing Kontsevich's homological mirror symmetry conjecture, it is likely that further study of the topological heterotic theory will have important applications to the study of (0,2)-mirror symmetry, a subject of considerable interest in both physics~\cite{Blumenhagen:1996vu, Blumenhagen:1997vt, Sharpe:1998wa, Adams:2003zy, Melnikov:2010sa, Melnikov:2012hk, Anderson:2016byt, Chen:2016tdd, Chen:2017mxp, Gu:2017nye, Gu:2019byn} and mathematics~\cite{Baraglia:2013wua, Garcia-Fernandez:2016ofz, Garcia-Fernandez:2018qcl, Clarke:2020erl, Alvarez-Consul:2020hbl,Alvarez-Consul:2023zon} over the past few decades.

\subsection*{Acknowledgements}
We thank Leron Borsten, Xenia de la Ossa, Mario Garcia-Fernandez, Sergei Gukov, Magdalena Larfors, Matthew Magill, George Smith, Daniel Waldram and Brian Williams for helpful discussions. AA is supported by NSF Grant No.~PHY2014195 and the Kadanoff Center for Theoretical Physics. AA also acknowledges the support of the European Union’s Horizon 2020 research and innovation program during the early stages of the project under the Marie Sklodowska-Curie grant agreement No.~838776. DDM is supported by the Norwegian Financial Mechanism 2014-2021 (project registration number 2019/34/H/ST1/00636). CSC is supported by an EPSRC New Investigator Award, grant number~EP/X014959/1. DT is supported by NSF Grant No.~PHYS-2112859. AA, CSC and DT gratefully acknowledge support from the workshop ``Supergravity, Generalized Geometry and Ricci Flow'' at the Simons Center for Geometry and Physics, Stony Brook University, at which some of the research for this paper was performed. The work of JJMI, EES and SW was partially supported by the University of Stavanger ISP grant No.~IN-12743, which also made possible a workshop in August 2022 during which some of the research for this work was performed.

\appendix

\section{Hodge diamond}
\label{app:Hodge}

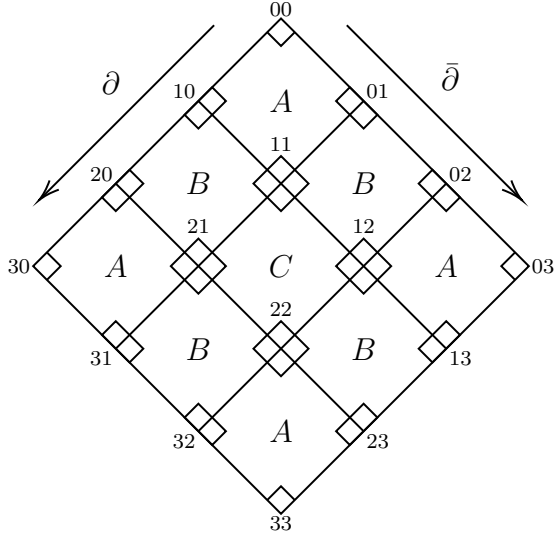
\begin{figure}
\centering

\tikzset{every picture/.style={line width=0.75pt}} %

\begin{tikzpicture}[x=0.75pt,y=0.75pt,yscale=-1,xscale=1]

\draw   (342,25) -- (467,150) -- (342,275) -- (217,150) -- cycle ;
\draw    (383.67,66.67) -- (258.67,191.67) ;
\draw    (425.33,108.33) -- (300.33,233.33) ;
\draw    (300.33,66.67) -- (425.33,191.67) ;
\draw    (258.67,108.33) -- (383.67,233.33) ;
\draw   (376.72,59.72) -- (390.61,73.61) -- (383.67,80.56) -- (369.78,66.67) -- cycle ;
\draw   (418.39,101.39) -- (432.28,115.28) -- (425.33,122.22) -- (411.44,108.33) -- cycle ;
\draw   (425.33,177.78) -- (432.28,184.72) -- (418.39,198.61) -- (411.44,191.67) -- cycle ;
\draw   (383.67,219.44) -- (390.61,226.39) -- (376.72,240.28) -- (369.78,233.33) -- cycle ;
\draw   (300.33,219.44) -- (314.22,233.33) -- (307.28,240.28) -- (293.39,226.39) -- cycle ;
\draw   (258.67,177.78) -- (272.56,191.67) -- (265.61,198.61) -- (251.72,184.72) -- cycle ;
\draw   (265.61,101.39) -- (272.56,108.33) -- (258.67,122.22) -- (251.72,115.28) -- cycle ;
\draw   (307.28,59.72) -- (314.22,66.67) -- (300.33,80.56) -- (293.39,73.61) -- cycle ;
\draw   (342,94.44) -- (355.89,108.33) -- (342,122.22) -- (328.11,108.33) -- cycle ;
\draw   (383.67,136.11) -- (397.56,150) -- (383.67,163.89) -- (369.78,150) -- cycle ;
\draw   (300.33,136.11) -- (314.22,150) -- (300.33,163.89) -- (286.44,150) -- cycle ;
\draw   (342,177.78) -- (355.89,191.67) -- (342,205.56) -- (328.11,191.67) -- cycle ;
\draw   (335.06,31.94) -- (342,38.89) -- (348.94,31.94) ;
\draw   (460.06,156.94) -- (453.11,150) -- (460.06,143.06) ;
\draw   (348.94,268.06) -- (342,261.11) -- (335.06,268.06) ;
\draw   (223.94,143.06) -- (230.89,150) -- (223.94,156.94) ;

\draw    (375.19,28.42) -- (462.16,115.4) ;
\draw [shift={(463.58,116.81)}, rotate = 225] [color={rgb, 255:red, 0; green, 0; blue, 0 }  ][line width=0.75]    (10.93,-3.29) .. controls (6.95,-1.4) and (3.31,-0.3) .. (0,0) .. controls (3.31,0.3) and (6.95,1.4) .. (10.93,3.29)   ;
\draw    (308.26,28.42) -- (221.29,115.4) ;
\draw [shift={(219.87,116.81)}, rotate = 315] [color={rgb, 255:red, 0; green, 0; blue, 0 }  ][line width=0.75]    (10.93,-3.29) .. controls (6.95,-1.4) and (3.31,-0.3) .. (0,0) .. controls (3.31,0.3) and (6.95,1.4) .. (10.93,3.29)   ;

\draw (421.24,46.32) node [anchor=north west][inner sep=0.75pt]    {$\bar{\partial }$};
\draw (250.21,50.29) node [anchor=north west][inner sep=0.75pt]    {$\partial $};

\draw (217 + 3* 250/6,25 + 1* 250/6) node [anchor=center][inner sep=0.75pt]    {$A$};
\draw (217 + 2* 250/6,25 + 2* 250/6) node [anchor=center][inner sep=0.75pt]    {$B$};
\draw (217 + 1* 250/6,25 + 3* 250/6) node [anchor=center][inner sep=0.75pt]    {$A$};
\draw (217 + 2* 250/6,25 + 4* 250/6) node [anchor=center][inner sep=0.75pt]    {$B$};
\draw (217 + 3* 250/6,25 + 5* 250/6) node [anchor=center][inner sep=0.75pt]    {$A$};
\draw (217 + 4* 250/6,25 + 4* 250/6) node [anchor=center][inner sep=0.75pt]    {$B$};
\draw (217 + 5* 250/6,25 + 3* 250/6) node [anchor=center][inner sep=0.75pt]    {$A$};
\draw (217 + 4* 250/6,25 + 2* 250/6) node [anchor=center][inner sep=0.75pt]    {$B$};
\draw (217 + 3* 250/6,25 + 3* 250/6) node [anchor=center][inner sep=0.75pt]    {$C$};

\draw (217 + 3* 250/6,25 + 0* 250/6) node [anchor=south][inner sep=0.75pt]    {$\scriptstyle 00$};
\draw (217 + 2* 250/6,25 + 1* 250/6) node [anchor=south east][inner sep=0.75pt]    {$\scriptstyle 10$};
\draw (217 + 1* 250/6,25 + 2* 250/6) node [anchor=south east][inner sep=0.75pt]    {$\scriptstyle 20$};
\draw (217 + 0* 250/6,25 + 3* 250/6) node [anchor=east][inner sep=0.75pt]    {$\scriptstyle 30$};
\draw (217 + 1* 250/6,25 + 4* 250/6) node [anchor=north east][inner sep=0.75pt]    {$\scriptstyle 31$};
\draw (217 + 2* 250/6,25 + 5* 250/6) node [anchor=north east][inner sep=0.75pt]    {$\scriptstyle 32$};
\draw (217 + 3* 250/6,25 + 6* 250/6) node [anchor=north][inner sep=0.75pt]    {$\scriptstyle 33$};
\draw (217 + 4* 250/6,25 + 5* 250/6) node [anchor=north west][inner sep=0.75pt]    {$\scriptstyle 23$};
\draw (217 + 5* 250/6,25 + 4* 250/6) node [anchor=north west][inner sep=0.75pt]    {$\scriptstyle 13$};
\draw (217 + 6* 250/6,25 + 3* 250/6) node [anchor=west][inner sep=0.75pt]    {$\scriptstyle 03$};
\draw (217 + 5* 250/6,25 + 2* 250/6) node [anchor=south west][inner sep=0.75pt]    {$\scriptstyle 02$};
\draw (217 + 4* 250/6,25 + 1* 250/6) node [anchor=south west][inner sep=0.75pt]    {$\scriptstyle 01$};
\draw (217 + 3* 250/6,25 -15 + 2* 250/6) node [anchor=south][inner sep=0.75pt]    {$\scriptstyle 11$};
\draw (217 + 3* 250/6,25 -15+ 4* 250/6) node [anchor=south][inner sep=0.75pt]    {$\scriptstyle 22$};
\draw (217 + 4* 250/6,25 -15+ 3* 250/6) node [anchor=south][inner sep=0.75pt]    {$\scriptstyle 12$};
\draw (217 + 2* 250/6,25 -15+ 3* 250/6) node [anchor=south][inner sep=0.75pt]    {$\scriptstyle 21$};

\end{tikzpicture}

\caption{Figure adapted from \cite{Pestun:2005rp}. Complex conjugation, Hodge duality and contraction with the holomorphic three-form $\Omega$ leave only three independent determinants. For example, $\det \Delta^{0,0}=A$ and $\det \Delta^{1,1}=AB^2C$.}
\label{fig: Hodge diamond}
\end{figure}

In the calculations of the partition function in the main text, we used certain properties of the Hodge diamond and the determinants on Calabi--Yau manifolds, first discussed by Pestun and Witten \cite{Pestun:2005rp}. In this appendix, we give a quick review of this material. When working on a K\"ahler threefold, the space of differential forms decomposes into subspaces $\Omega^{p,q}$, where $p,q \in \{ 0,1,2,3\} $. On a three-fold, this leads to nine possible subspaces. Since these manifolds also come equipped with the operators $\del $ and $\bar{\del}$, these spaces fit into a double complex known as the Hodge diamond, illustrated in Figure \ref{fig: Hodge diamond}. We see that $\del $ and $\bar{\del}$ act along diagonal edges, and their respective adjoint operators, $\del^\dagger $ and $\bar{\del}^\dagger$, act in the opposite directions. Each of the nine diamond faces are labelled by $A$, $B$ or $C$. As we will see, these can be associated to certain products of determinants of the Laplacian restricted to $(p,q)$-forms.

\begin{figure}
        \centering
        \begin{tikzpicture}[scale=1.5]
            \draw[black] (-0.5,-0.5) -- (0.5,0.5);
            \draw[black] (-0.5,0.5) -- (0.5,-0.5);
            \draw[black] (-0.25,0) -- (0,0.25);
            \draw[black] (0,0.25) -- (0.25,0);
            \draw[black] (0.25,0) -- (0,-0.25);
            \draw[black] (0,-0.25) -- (-0.25,0);
            \node[circle,align=center,text=black,inner sep=0pt,minimum size=0pt] at (-0.5,0) {$\OmAC$};
            \node[circle,align=center,text=black,inner sep=0pt,minimum size=0pt] at (0.5,0) {$\OmBD$};
            \node[circle,align=center,text=black,inner sep=0pt,minimum size=0pt] at (0,0.5) {$\OmAB{}$};
            \node[circle,align=center,text=black,inner sep=0pt,minimum size=0pt] at (0,-0.5) {$\OmCD$};
        \end{tikzpicture}
        \label{fig:Dolbeault_vertex}
    \hfill
    \caption{Subspaces at each vertex.}
\end{figure}
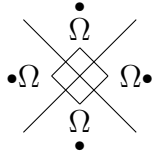

In Figure \ref{fig: Hodge diamond}, we have also drawn smaller squares at each of the vertices. These are to depict the decomposition of the space of differential forms at each vertex, following from
\begin{equation}\label{eq:subspaces}
    \Omega = H \oplus \OmAB{} \oplus \OmCD \oplus \OmAC{} \oplus \OmBD\, .
\end{equation}
Here $H$ denotes the set of harmonic forms (which we have ignored in our computations), while the spaces $ \OmAB$, $\OmCD$, $\OmAC$, and $\OmBD$ are orthogonal subspaces of $(p,q)$-forms which are in the image of certain combinations of $\del$, $\del^\dagger$ and their conjugates. For instance, for $(1,1)$-forms, the subspace $\OmAB{}^{1,1}$ is the image of $\del \Bar{\del}$ acting on $\Omega^{0,0}$. The subspace $\OmBD{}^{1,1}$ is given by the image of $\del \bar{\del}^\dagger$ acting on $\Omega^{0,2}$, and so on. We represent the subspaces by the small squares associated with each vertex in the Hodge diamond, as depicted in Figure \ref{fig:Dolbeault_vertex}.\footnote{As we have mentioned, we ignore harmonic forms, but for completeness we can imagine that the harmonic forms live at centre of each vertex. If these are reintroduced, the determinants that appear in this appendix should be properly denoted by $\det{}'$, indicating that the zero-modes are removed.} To each vertex and face, one can associate certain determinants of Laplacians. Since $\Delta$ commutes with the relevant operators, the Laplacian on $(p,q)$-forms can also be decomposed as in \eqref{eq:subspaces}.

Due to symmetry under complex conjugation and the Hodge star, not all of the face and vertex Laplacians are independent. Furthermore, if the threefold is Calabi--Yau, the Laplacian commutes with multiplication by the holomorphic three-form $\Omega$. Taken together, only three of the face Laplacians in the Hodge diamond are independent. The three independent faces are the four corner faces, the four edge-centred faces, and the central face. These Laplacians are denoted by $\Delta_A$, $\Delta_B$ and $\Delta_C$, with their respective determinants $A$, $B$, and $C$ shown on Figure \ref{fig: Hodge diamond}. Finally, the determinants of the vertex Laplacians (i.e.\ those acting on standard $(p,q)$-forms) can be expressed in terms of $A$, $B$ and $C$ as
\begin{align}
\det\Delta_{00} &= \det \Delta_{33} = \det \Delta_{30} = \det \Delta_{03} = A\, ,\\
\begin{split}
    \det \Delta_{10} &= \det \Delta_{01} = \det \Delta_{32} = \det \Delta_{23} = \det \Delta_{20} = \det \Delta_{02} \\
        &= \det \Delta_{13} = \det \Delta_{31} = AB\, ,
\end{split}\\
\det \Delta_{11} &= \det \Delta_{21} = \det \Delta_{12} = \det \Delta_{22} = AB^2C\, .
\end{align}

\section{The heterotic differential}
\label{app:FluxOperator}
In this appendix we describe the heterotic twisted differential $\bar D$ on $Q=T^{1,0}\oplus\End(V)\oplus  T^{*1,0}$ which has appeared in \cite{Gualtieri:2010fd, Anderson:2014xha, delaOssa:2014cia, delaOssa:2015maa, Ashmore:2018ybe, McOrist:2021dnd}. We will follow the approach taken in those works and neglect the subtleties regarding the extra connection on $\End(T)$ appearing in the heterotic Bianchi identity. The operator (including $\alpha'$ to get the correct heterotic Bianchi identity) reads
\begin{equation}
    \bar D= \begin{pmatrix} 
         \bar\partial & \mathcal{F}^* & \tilde{H} \\
         0 & \bar\partial_{A} & \mathcal{F} \\
         0 & 0 & \bar\partial
    \end{pmatrix}\;\;\;
    \begin{matrix} 
         T^{*(1,0)}X \\
         \End(V) \\
         T^{(1,0)}X 
    \end{matrix}\;\;,
\end{equation}
where the maps $\mathcal{F}$, $\mathcal{F}^*$ and $\tilde{H}$ are given by
\begin{equation}
\arraycolsep = 1.4pt
\begin{array}{rcrcl}
    \mu\in\Omega^{(0,p)}(T^{*(1,0)}X) \colon & \qquad & \mathcal{F}(\mu) &=& F_{a\bar b}\,\dd z^{\bar b}\wedge\mu^a\,, \qquad   \tilde{H}_b(\mu)= \mu^a\wedge\tilde{H}_{ab\bar c}\dd z^{\bar c}\\[3pt]
    \alpha\in\Omega^{(0,p)}(\End(V))\colon & &  \mathcal{F}^*_b(\alpha) &=& \frac{\alpha'}{4} \tr ( F_{b\bar c}\,\dd z^{\bar c}\wedge \alpha)\,,
\end{array}
\end{equation}
where $F$ is the curvature of the holomorphic gauge connection, and $\tilde{H}$ is a $\partial$-closed $(2,1)$-form. The $\partial$-closure is required for the higher order $L_\infty$-structure to work, which will not be covered here (see \cite{Ashmore:2018ybe} for further details). For on-shell heterotic Hull--Strominger solutions, $\tilde{H}=H^{(2,1)}=\ii\partial\omega$, and requiring that the operator $\bar D$ is nilpotent leads to the heterotic Bianchi identity
\begin{equation}
    \dd H=-2\ii\partial\bar\partial\omega=\frac{\alpha'}{4}\tr(F\wedge F)\,.
\end{equation}

\subsection{Curvature polynomials on \texorpdfstring{$Q$}{Q}}
\label{app:CurvPolQ}
As is relevant for the current paper, the operator $\bar D$ forms part of a Chern-connection $D_c$ on $Q$. We write
\begin{equation}
    D_c=\bar D + D\,.
\end{equation}
Let us compute the curvature of $D_c$. This is most compactly done by using the metric to combine the holomorphic tangent and cotangent bundles into the complexified tangent bundle
\begin{equation}
    T_{\mathbb{C}}\cong T^{*1,0}\oplus T^{1,0}\,,
\end{equation}
on which the Chern connection takes the form \cite{delaOssa:2017gjq, Garcia-Fernandez:2023nil, Garcia-Fernandez:2023vah}
\begin{equation}
    D_c=\begin{pmatrix} 
         \dd_A & \mathcal{F} \\
         \mathcal{F}^* & \dd_{\nabla^-}
    \end{pmatrix}\,,
\end{equation}
where $d_A$ is the gauge connection and $\dd_{\nabla^-}$ is the Hull connection \cite{Hull:1986kz}, given by
\begin{equation}
    \nabla^{\pm}=\nabla^{LC}\pm\tfrac{1}{2}H\,,
\end{equation}
where $\nabla^+$ is the supersymmetry preserving Bismut connection. As noted in \cite{delaOssa:2017gjq, Garcia-Fernandez:2023nil, Garcia-Fernandez:2023vah}, the full Hull--Strominger system requires $D_c$ to be hermitian Yang--Mills, though the F-term constraints coming from the superpotential only impose holomorphy. The curvature is then given by
\begin{equation}
    D_c^2=\begin{pmatrix} 
         F+\mathcal{F}\mathcal{F}^* & \dd_{A,\nabla^-}\mathcal{F} \\
         \dd_{A,\nabla^-}\mathcal{F}^* & R_{\nabla^-}+\mathcal{F}^*\mathcal{F}
    \end{pmatrix}\,.
\end{equation}
Note that $\mathcal{F}^*\sim\mathcal{O}(\alpha')$. Hence, $D_c$ becomes upper-triangular at zeroth order. Furthermore, assuming we are on a Calabi--Yau background at zeroth order in $\alpha'$, $R_{\nabla^-}=R+\mathcal{O}(\alpha')$ where $R$ is the curvature tensor of the Calabi--Yau background metric.  Modulo higher order corrections, any curvature polynomial in $D_c$ is thus also upper-triangular. Hence, if we restrict to a diagonal metric on $Q$, modulo higher orders the metric anomaly therefore becomes the same as using the diagonal Chern connection on $Q$, where the tangent bundle connection is given by the Chern connection of the Calabi--Yau metric.

\section{Some group theory}

\subsection{Trace identities for some groups}\label{app:trace_identities}

\begin{align}
    SO(8)\colon     &  & \Tr(T^{2}) & = 6\tr (T^{2})\, , \\
    &  & \Tr(T^{4}) & = 3 \tr(T^{2})^{2}\, , \nonumber\\[3pt]
	SU(3)\colon &  & \tr(T^{2}) & =\tfrac{1}{2}\,,\\
	&  & \tr(T^{4}) & =\tfrac{1}{2}\tr(T^{2})^2\,,\nonumber\\
	&  & \Tr(T^{2}) & =6\tr(T^{2})\,,\nonumber\\
	&  & \Tr(T^{4}) & =9\tr(T^{2})^{2}\,,\nonumber\\[3pt]
	SU(5)\colon &  & \tr(T^{2}) & =\tfrac{1}{2}\,,\\
	&  & \Tr(T^{2}) & =10\tr(T^{2})\,,\nonumber\\
	&  & \Tr(T^{4}) & =10\tr(T^{4})+6\tr(T^{2})^{2}\,,\nonumber\\[3pt]
	G_{2}\colon &  & \tr(T^{2}) & =1\,,\\
	&  & \tr(T^{4}) & =\tfrac{1}{4}\tr(T^{2})^{2}\,,\nonumber\\
	&  & \Tr(T^{2}) & =4\tr(T^{2})\,,\nonumber\\
	&  & \Tr(T^{4}) & =\tfrac{5}{2}\tr(T^{2})^{2}\,,\nonumber\\[3pt]
	F_{4}\colon &  & \tr(T^{2}) & =3\,,\\
	&  & \tr(T^{4}) & =\tfrac{1}{12}\tr(T^{2})^{2}\,,\nonumber\\
	&  & \Tr(T^{2}) & =3\tr(T^{2})\,,\nonumber\\
	&  & \Tr(T^{4}) & =\tfrac{5}{12}\tr(T^{2})^{2}\,,\nonumber\\[3pt]
	E_{6}\colon &  & \tr(T^{2}) & =3\,,\\
	&  & \tr(T^{4}) & =\tfrac{1}{12}\tr(T^{2})^{2}\,,\nonumber\\
	&  & \Tr(T^{2}) & =4\tr(T^{2})\,,\nonumber\\
	&  & \Tr(T^{4}) & =\tfrac{1}{2}\tr(T^{2})^{2}\,,\nonumber\\[3pt]
	E_{7}\colon &  & \tr(T^{2}) & =6\,,\\
	&  & \tr(T^{4}) & =\tfrac{1}{24}\tr(T^{2})^{2}\,,\nonumber\\
	&  & \Tr(T^{2}) & =3\tr(T^{2})\,,\nonumber\\
	&  & \Tr(T^{4}) & =\tfrac{1}{6}\tr(T^{2})^{2}\,,\nonumber\\[3pt]
	E_{8}\colon &  & \Tr(T^{2}) & =30\,,\\
	&  & \Tr(T^{4}) & =\tfrac{1}{100}\Tr(T^{2})^{2}\,.\nonumber
\end{align}

\subsection{Anomaly polynomial factorisation}
\label{app:AnomalyPolynomials}

Here are the factorised anomaly polynomials for the gauge groups $G_{1}\times G_{2}$ using \eqref{eq:AnomalyPolGauge2}.

\begin{align}
    SU(3)\times SU(3)\colon& \qquad \int_{X}\tfrac{1}{48}(\ch_{2}(X)-6\ch_{2}(V_{1}))^{2} + \tfrac{1}{48}(\ch_{2}(X)-6\ch_{2}(V_{2}))^{2}\, ,\\[3pt]
    SU(3)\times G_{2}\colon& \qquad \int_{X}\tfrac{1}{48}(\ch_{2}(X)-6\ch_{2}(V_{1}))^{2} + \tfrac{1}{120}(2\ch_{2}(X)-5\ch_{2}(V_{2}))^{2}\, ,\\[3pt]
    SU(3)\times SO(8)\colon& \qquad \int_{X}\tfrac{1}{48}(\ch_{2}(X)-6\ch_{2}(V_{1}))^{2} + \tfrac{1}{16}(\ch_{2}(X)-2\ch_{2}(V_{2}))^{2}\, ,\\[3pt]
    SU(3)\times F_{4}\colon& \qquad \int_{X}\tfrac{1}{48}(\ch_{2}(X)-6\ch_{2}(V_{1}))^{2} + \tfrac{1}{720}(9\ch_{2}(X)-5\ch_{2}(V_{2}))^{2}\, ,\\[3pt]
    SU(3)\times E_{6}\colon& \qquad \int_{X}\tfrac{1}{48}(\ch_{2}(X)-6\ch_{2}(V_{1}))^{2} + \tfrac{1}{24}(2\ch_{2}(X)-\ch_{2}(V_{2}))^{2}\, ,\\[3pt]
    SU(3)\times E_{6}\colon& \qquad \int_{X}\tfrac{1}{48}(\ch_{2}(X)-6\ch_{2}(V_{1}))^{2} + \tfrac{1}{24}(2\ch_{2}(X)-\ch_{2}(V_{2}))^{2}\, ,\\[3pt]
    SU(3)\times E_{7}\colon& \qquad \int_{X}\tfrac{1}{48}(\ch_{2}(X)-6\ch_{2}(V_{1}))^{2} + \tfrac{1}{288}(9\ch_{2}(X)-2\ch_{2}(V_{2}))^{2}\, ,\\[3pt]
    SU(3)\times E_{8}\colon& \qquad \int_{X}\tfrac{1}{48}(\ch_{2}(X)-6\ch_{2}(V_{1}))^{2} + \tfrac{1}{1200}(25\ch_{2}(X)-\ch_{2}(V_{2}))^{2}\, ,\\[3pt]
    G_{2}\times G_{2}\colon& \qquad \int_{X}\tfrac{1}{120}(2\ch_{2}(X)-5\ch_{2}(V_{1}))^{2} + \tfrac{1}{120}(2\ch_{2}(X)-5\ch_{2}(V_{2}))^{2}\, ,\\[3pt]
    G_{2}\times SO(8)\colon& \qquad \int_{X}\tfrac{1}{120}(2\ch_{2}(X)-5\ch_{2}(V_{1}))^{2} + \tfrac{1}{16}(\ch_{2}(X)-2\ch_{2}(V_{2}))^{2}\, ,\\[3pt]
    G_{2}\times F_{4}\colon& \qquad \int_{X}\tfrac{1}{120}(2\ch_{2}(X)-5\ch_{2}(V_{1}))^{2} + \tfrac{1}{720}(9\ch_{2}(X)-5\ch_{2}(V_{2}))^{2}\, ,\\[3pt]
    G_{2}\times E_{6}\colon& \qquad \int_{X}\tfrac{1}{120}(2\ch_{2}(X)-5\ch_{2}(V_{1}))^{2} + \tfrac{1}{24}(2\ch_{2}(X)-\ch_{2}(V_{2}))^{2}\, ,\\[3pt]
    G_{2}\times E_{7}\colon& \qquad \int_{X}\tfrac{1}{120}(2\ch_{2}(X)-5\ch_{2}(V_{1}))^{2} + \tfrac{1}{288}(9\ch_{2}(X)-2\ch_{2}(V_{2}))^{2}\, ,\\[3pt]
    G_{2}\times E_{8}\colon& \qquad \int_{X}\tfrac{1}{120}(2\ch_{2}(X)-5\ch_{2}(V_{1}))^{2} + \tfrac{1}{1200}(25\ch_{2}(X)-\ch_{2}(V_{2}))^{2}\, ,\\[3pt]
    SO(8)\times SO(8)\colon& \qquad \int_{X}\tfrac{1}{16}(\ch_{2}(X)-2\ch_{2}(V_{1}))^{2} + \tfrac{1}{16}(\ch_{2}(X)-2\ch_{2}(V_{2}))^{2}\, ,\\[3pt]
    SO(8)\times F_{4}\colon& \qquad \int_{X}\tfrac{1}{16}(\ch_{2}(X)-2\ch_{2}(V_{1}))^{2} + \tfrac{1}{720}(9\ch_{2}(X)-5\ch_{2}(V_{2}))^{2}\, ,\\[3pt]
    SO(8)\times E_{6}\colon& \qquad \int_{X}\tfrac{1}{16}(\ch_{2}(X)-2\ch_{2}(V_{1}))^{2} + \tfrac{1}{24}(2\ch_{2}(X)-\ch_{2}(V_{2}))^{2}\, ,\\[3pt]
    SO(8)\times E_{7}\colon& \qquad \int_{X}\tfrac{1}{16}(\ch_{2}(X)-2\ch_{2}(V_{1}))^{2} + \tfrac{1}{288}(9\ch_{2}(X)-2\ch_{2}(V_{2}))^{2}\, ,\\[3pt]
    SO(8)\times E_{8}\colon& \qquad \int_{X}\tfrac{1}{16}(\ch_{2}(X)-2\ch_{2}(V_{1}))^{2} + \tfrac{1}{1200}(25\ch_{2}(X)-\ch_{2}(V_{2}))^{2}\, ,\\[3pt]
    F_{4}\times F_{4}\colon& \qquad \int_{X}\tfrac{1}{720}(9\ch_{2}(X)-5\ch_{2}(V_{1}))^{2} + \tfrac{1}{720}(9\ch_{2}(X)-5\ch_{2}(V_{2}))^{2}\, ,\\[3pt]
    F_{4}\times E_{6}\colon& \qquad \int_{X}\tfrac{1}{720}(9\ch_{2}(X)-5\ch_{2}(V_{1}))^{2} + \tfrac{1}{24}(2\ch_{2}(X)-\ch_{2}(V_{2}))^{2}\, ,\\[3pt]
    F_{4}\times E_{7}\colon& \qquad \int_{X}\tfrac{1}{720}(9\ch_{2}(X)-5\ch_{2}(V_{1}))^{2} + \tfrac{1}{288}(9\ch_{2}(X)-2\ch_{2}(V_{2}))^{2}\, ,\\[3pt]
    F_{4}\times E_{8}\colon& \qquad \int_{X}\tfrac{1}{720}(9\ch_{2}(X)-5\ch_{2}(V_{1}))^{2} + \tfrac{1}{1200}(25\ch_{2}(X)-\ch_{2}(V_{2}))^{2}\, ,\\[3pt]
    E_{6}\times E_{6}\colon& \qquad \int_{X}\tfrac{1}{24}(2\ch_{2}(X)-\ch_{2}(V_{1}))^{2} + \tfrac{1}{24}(2\ch_{2}(X)-\ch_{2}(V_{2}))^{2}  \, ,\\[3pt]
    E_{6}\times E_{7}\colon& \qquad \int_{X}\tfrac{1}{24}(2\ch_{2}(X)-\ch_{2}(V_{1}))^{2} + \tfrac{1}{288}(9\ch_{2}(X)-2\ch_{2}(V_{2}))^{2}\, ,\\[3pt]
    E_{6}\times E_{8}\colon& \qquad \int_{X}\tfrac{1}{24}(2\ch_{2}(X)-\ch_{2}(V_{1}))^{2} + \tfrac{1}{1200}(25\ch_{2}(X)-\ch_{2}(V_{2}))^{2}\, ,\\[3pt]
    E_{7}\times E_{7}\colon& \qquad \int_{X}\tfrac{1}{288}(9\ch_{2}(X)-2\ch_{2}(V_{1}))^{2} + \tfrac{1}{288}(9\ch_{2}(X)-2\ch_{2}(V_{2}))^{2}\, ,\\[3pt]
    E_{7}\times E_{8}\colon& \qquad \int_{X}\tfrac{1}{288}(9\ch_{2}(X)-2\ch_{2}(V_{1}))^{2} + \tfrac{1}{1200}(25\ch_{2}(X)-\ch_{2}(V_{2}))^{2}\, ,\\[3pt]
    E_{8}\times E_{8}\colon& \qquad \int_{X}\tfrac{1}{1200}(25\ch_{2}(X)-\ch_{2}(V_{1}))^{2} + \tfrac{1}{1200}(25\ch_{2}(X)- \ch_{2}(V_{2}))^{2}\, .
\end{align}

\bibliographystyle{utphys}
\bibliography{Refs}

\end{document}